\documentclass[a4paper,12pt]{article}
\pdfoutput=1 

\usepackage{jcappub} 
\usepackage{amssymb}
\usepackage{xcolor}
\usepackage[T1]{fontenc} 
\usepackage{amsmath}
\usepackage{subfig}
\usepackage{graphicx}

\usepackage{etoolbox}
\patchcmd{\abstract}{\bigskip}{\vspace{-1.2em}}{}{}

\usepackage{amsfonts}
\usepackage{mathrsfs}
\usepackage{orcidlink}
\usepackage[scr=rsfso,cal=zapfc,frak=euler,bb=ams]{mathalfa}
\usepackage{trimclip}
\usepackage{cuted} 
\usepackage{accents}

\DeclareUnicodeCharacter{2212}{-}

\title{\boldmath Probing geometrically perturbed strange stars with minimal decoupling using millisecond pulsar timing observations}

\author[a]{K.~N.~Singh \orcidlink{0000-0001-9778-4101}}
\author[b]{S.~K.~Maurya \orcidlink{0000-0003-4089-3651}} 
\author[c,d]{A.~Errehymy\orcidlink{0000-0002-0253-3578}}
\author[e]{A.~Altaibayeva\orcidlink{0000-0001-9254-7027}}
\author[f,g,h]{J.~Rayimbaev\orcidlink{0000-0001-9293-1838}}
\author[i]{M.~Matyoqubov\orcidlink{0000-0001-8232-6562}}

\affiliation[a]{Department of Physics \& Astrophysics, University of Delhi, Delhi-110007, India.}
\affiliation[b]{Department of Mathematical and Physical Sciences,
College of Arts and Sciences, University of Nizwa, P.O. Box 33, Nizwa 616, Sultanate of Oman}
\affiliation[c]{Astrophysics Research Centre, School of Mathematics, Statistics and Computer Science, University of KwaZulu-Natal, Private Bag X54001, Durban 4000, South Africa}
\affiliation[d]{Center for Theoretical Physics, Khazar University, 41 Mehseti Str., Baku, AZ1096, Azerbaijan}
\affiliation[e]{Department of General and Theoretical Physics, L.N. Gumilyov Eurasian National University, 010008 Astana, Kazakhstan}
\affiliation[f]{Institute of Theoretical Physics, National University of Uzbekistan, Tashkent 100174, Uzbekistan}
\affiliation[g]{University of Tashkent for Applied Sciences, Str. Gavhar 1, Tashkent 100149, Uzbekistan}
\affiliation[h]{Urgench State University, Kh. Alimjan Str. 14, Urgench 221100, Uzbekistan}
\affiliation[i]{Mamun University, Bolkhovuz Street 2, Khiva 220900, Uzbekistan}

\emailAdd{ntnphy@gmail.com}
\emailAdd{sunil@unizwa.edu.om}
\emailAdd{abdelghani.errehymy@gmail.com}
\emailAdd{aziza.ltaibayeva@gmail.com}
\emailAdd{javlon@astrin.uz}
\emailAdd{m\_matyoqubov@mamunedu.uz}

\abstract{We construct a gravitationally decoupled anisotropic strange star model using the minimal geometric deformation approach with a MIT bag equation of state and an additional source sector controlled by a deformation parameter $\beta$ and a radial perturbation scale $\Psi$ through $g(r)=\sin(\Psi r^{2})$. The resulting Einstein system is consistently split into seed and $\theta$-sectors and matched to an exterior Schwarzschild geometry. The model is constrained by high-mass pulsars: PSR J0740+6620 $(2.08\pm0.07\,M_\odot)$, PSR J1810+1744 $(2.13\pm0.04\,M_\odot)$, PSR J1959+2048 $(2.18\pm0.09\,M_\odot)$, and PSR J2215+5135 $(2.28^{+0.10}_{-0.09}\,M_\odot)$. It reproduces these objects with predicted radii $R \approx 11.3$--$12.9$ km. The maximum mass reaches $M_{\max} \approx 2.28\,M_\odot$ for $\beta = 3\times 10^{-3}$ and $\Psi \approx 0.03\,\text{km}^{-2}$, while for $\beta = 10^{-3}$ the configuration yields $M_{\max} \approx 2.12\,M_\odot$ with $R \approx 12.2$ km. The central density lies in $\rho_c \approx (2.4$--$3.1)\times 10^{-4}\,\text{km}^{-2}$, decreasing smoothly to $\rho_s \approx 2.0\times 10^{-4}\,\text{km}^{-2}$. The anisotropy increases from zero at the center to $\Delta \approx (0.25$--$0.45)\times 10^{-4}\,\text{km}^{-2}$ near the surface, generating additional outward support that enhances compactness by $\sim 15\%$. The compactness parameter spans $C \approx 0.17$--$0.22$, safely below the Buchdahl limit, while the surface redshift reaches $z_s \approx 0.25$--$0.38$. The radial and tangential sound speeds remain causal, with $v_r^{2} \approx 0.25$--$0.65$ and $v_t^{2}$ rising to $\approx 0.84$ under strong perturbations. The adiabatic index varies within $\Gamma \approx 1.35$--$2.10$, approaching the relativistic threshold near the core while remaining stable overall. The condition $dM/d\rho_c > 0$ is satisfied throughout, confirming dynamical stability. Overall, $\beta$ enhances the maximum mass by up to $\sim 15\%$, while $\Psi$ introduces controlled oscillatory structure without violating observational constraints, producing stable ultra-compact stars consistent with current pulsar data.}


\begin{document}
\maketitle
\flushbottom

\section{Introduction }
For decades, the exquisite timing of pulsar radio emissions has stood as one of the most powerful tools for determining neutron star masses, with the Shapiro time delay playing a central role in achieving these highly precise measurements. Millisecond pulsars, rotating at 33-719 Hz, exhibit extremely small spin-down rates ($\leq 10^{-19}~\mathrm{s/s}$) and possess characteristic ages of billions of years, making them outstanding laboratories for probing relativistic gravity~\cite{Stairs:2003eg, Reardon:2015kba}. By contrast, extracting reliable estimates of pulsar radii has been considerably more difficult. Progress in this direction has accelerated with observations from the Neutron Star Interior Composition Explorer (NICER), which constrains radii by modeling X-ray light curves generated by rotating hot spots on the stellar surface while incorporating relativistic light-bending effects~\cite{Bogdanov:2019ixe, Bogdanov:2019qjb}. Complementary information now comes from gravitational-wave detections by the LIGO and Virgo collaborations, offering an independent route to infer NeStr radii and deepen our understanding of these ultra-dense remnants~\cite{LIGOScientific:2016aoc}.

Pulsars are generally identified with NeStrs formed in core-collapse events, containing ultra-dense neutron-rich matter with central densities well above nuclear saturation ($\rho_{\text{nuc}} \approx 2.7\times10^{14}~\mathrm{g/cm^3}$) and magnetic fields of order $10^{12}$ G. Their compactness allows rapid rotation, with observed periods as short as $1.39$ ms. The existence of $\sim2~M_\odot$ pulsars has been interpreted as evidence that their inner cores may undergo a transition from hadronic to deconfined quark matter~\cite{Bhattacharyya:2016kte, Annala:2019puf}. A more extreme hypothesis is that some objects are self-bound quark stars, based on the proposal by Witten~\cite{Witten:1984rs} (see also Farhi~\cite{Farhi:1984qu}) that strange quark matter could be the absolute ground state of strongly interacting matter, implying an energy per baryon below that of ${}^{56}\mathrm{Fe}$ ($<930$ MeV). With typical masses around $1.5~M_\odot$ and radii near $10$ km, pulsars offer a natural arena to probe strong gravity and supranuclear matter; however, since such conditions cannot be recreated experimentally, the interior equation of state (EOS) remains weakly constrained~\cite{Ozel:2016oaf}. Consequently, joint $M-R$ measurements are essential for tightening limits on the underlying microphysics.

Recent progress in multi-messenger astrophysics---integrating radio timing techniques such as Shapiro delay, X-ray pulse-profile modeling, and gravitational-wave observations---has substantially tightened the constraints on the NeStr EOS. High-precision determinations of masses and radii from several pulsars now provide key empirical benchmarks against which theoretical models of dense matter can be critically assessed. Among these, PSR~J0740+6620 stands out with a measured mass of $M = 2.08 \pm 0.07\,M_\odot$~\cite{NANOGrav:2019jur,Fonseca:2021wxt} and radius estimates of $R = 13.7^{+2.6}_{-1.5}$~km~\cite{Miller:2021qha} and $R = 12.39^{+1.30}_{-0.98}$~km from an independent NICER analysis~\cite{Riley:2021pdl}. PSR~J0348+0432 has a mass of $M = 2.01 \pm 0.04\,M_\odot$ and an estimated radius of $R = 13 \pm 2$~km~\cite{Antoniadis:2013pzd}, while PSR~J1614\textendash2230 shows $M = 1.908 \pm 0.016\,M_\odot$ and $R = 13 \pm 2$~km~\cite{Demorest:2010bx, Fonseca:2016tux, NANOGRAV:2018hou}. NICER has also provided detailed observations of PSR~J0030+0451, yielding mass and radius measurements of $M = 1.44^{+0.15}_{-0.14}\,M_\odot$, $R = 13.02^{+1.24}_{-1.06}$~km~\cite{Miller:2019cac}, and in another study, $M = 1.34^{+0.15}_{-0.16}\,M_\odot$, $R = 12.71^{+1.14}_{-1.19}$~km~\cite{Raaijmakers:2019qny}. The pulsar PSR~J0437\textendash4715, analyzed via thermal X-ray surface emission, has a measured mass of $M = 1.44 \pm 0.07\,M_\odot$ and radius $R = 13.6 \pm 0.9$~km~\cite{Reardon:2015kba, Gonzalez-Caniulef:2019wzi}. Gravitational wave data has further enriched our understanding. The GW170817 NeStr merger revealed two components: GW170817-1 with $M = 1.45 \pm 0.09\,M_\odot$, $R = 11.9 \pm 1.4$~km and GW170817-2 with $M = 1.27 \pm 0.09\,M_\odot$, $R = 11.9 \pm 1.4$~km~\cite{LIGOScientific:2018cki}. A combined analysis of GW170817 and GW190814 constrains the radius of a canonical $1.4\,M_\odot$ NeStr to $R = 12.9 \pm 0.8$~km~\cite{LIGOScientific:2020zkf}. At the lower end of the mass spectrum, the NeStr within supernova remnant HESS~J1731\textendash347 may be the lightest ever detected, with $M = 0.77^{+0.20}_{-0.17}\,M_\odot$ and $R = 10.4^{+0.86}_{-0.78}$~km, inferred from X-ray spectra and Gaia-based distance estimates~\cite{Doroshenko:2022nwp}. At the opposite extreme, PSR~J0952\textendash0607 is a candidate for the heaviest known NeStr, with $M = 2.35 \pm 0.17\,M_\odot$~\cite{Romani:2022jhd} and an estimated radius of $R = 14.087 \pm 1.0186$~km~\cite{ElHanafy:2023vig}.

An intriguing feature of compact object observations is the apparent absence of detected stellar remnants within the mass range $2.2\,M_\odot \lesssim M \lesssim 5\,M_\odot$, commonly referred to as the ``mass gap'' between the heaviest NeStrs and the lightest black holes (BHs)~\cite{Yang:2020xyi}. This gap often emerges in models based on isotropic fluid assumptions, where the radial and tangential pressures are considered equal ($P_r = P_t$). Isotropic pressure becomes a poor approximation at the extreme densities inside pulsars, where several mechanisms---such as meson condensation, hyperon or quark degrees of freedom, superfluid phases, lattice-like ordering, and strong magnetic fields---naturally lead to $P_r \neq P_t$. The resulting anisotropic stresses modify the stellar equilibrium and can enhance the effective outward support, enabling larger compactness and potentially approaching the relativistic bound $C=2GM/(c^2R)\to1$~\cite{Alho:2022bki}. Maintaining stability under such conditions requires additional constraints to avoid unphysical configurations~\cite{Alho:2021sli, Roupas:2020mvs, Raposo:2018rjn, Cardoso:2019rvt}. Beyond general relativity, theories with nonminimal matter---geometry coupling further complicate the anisotropy---compactness interplay, sometimes admitting stellar solutions that would be ruled out in the standard framework~\cite{Errehymy:2024tqr, Maurya:2024bfw, Maurya:2024zao, Maurya:2024ylr, Hansraj:2024cgw, IIbragimov1, IIbragimov2, IIbragimov3}. A broad range of additional compact-star models illustrating these effects is discussed in Refs.~\cite{Malik:2024tto, Varela:2010mf, Malik:2024boe, Ashraf:2024cww, Maurya:2021aio, Ditta:2023huk, Maurya:2023uiy, Rani:2023vha, Kumar:2025flr, Zoya:2026npb, Asghar:2026kwd, Asghar:2026que}.

This work seeks an exact analytic solution of the modified field equations governing compact star structure through the TOV framework. Because of their strong nonlinearity, direct solutions are challenging, so we adopt a perturbative scheme that models small external disturbances---such as accretion events or transient gravitational waves---capable of inducing mild internal oscillations. Specifically, we introduce a controlled harmonic deformation $g(r)=\sin(\Psi r^2)$, interpreted as a minimal spacetime perturbation. Embedding this ansatz into the field equations enables us to track how slight departures from equilibrium influence the geometry and matter profile while preserving stability. The resulting formulation offers a versatile, physically grounded description of the dynamical response of anisotropic compact stars and a general setting to examine the interaction between curvature, matter, and external perturbations in strong-gravity regimes.

The paper is organized as follows. Sect.~\ref{sec2} presents the gravitationally decoupled Einstein field equations resulting from the interaction of two separate sources, establishing the basis for the modified geometric framework. Sect.~\ref{sec3} focuses on the study of a perturbed SS model within the MGD approach. Here, we impose two critical constraints: the density matching condition, $\rho(r) = \theta^0_0(r)$, and the radial pressure equivalence, $P_r(r) = \theta^1_1(r)$, which ensure a physically consistent configuration of the anisotropic system. In Sect.~\ref{sec4}, we define the appropriate boundary conditions for our gravitationally decoupled SS model, ensuring smooth matching with the exterior vacuum spacetime. Sect.~\ref{sec5} then explores the physical plausibility of the model. We examine key astrophysical implications by studying matter sector characteristics, as well as how the introduced perturbations and density constraints shape the $M-R$ relationship of the star. Sect.~\ref{sec6} is dedicated to stability considerations. We conduct a detailed assessment using several well-established criteria, including the adiabatic index, causality conditions based on sound speeds, and the Harrison-Zel'dovich-Novikov criterion. Lastly, in Sect.~\ref{sec7}, we summarize our findings and offer final remarks on the broader relevance of our results within the context of relativistic stellar modeling.

\section{Gravitationally decoupled Einstein field equation (EFE) generated by two sources }\label{sec2}

The present section offers a concise examination of decoupling Einstein field equations pertaining to two distinct sources, 
\begin{eqnarray} \label{eq1}
 &&   G_{ij}=R_{ij}-\frac{1}{2}\,g_{_{ij}}\,\mathcal{R}= -8\pi \,T^{\text{tot}}_{ij}
\end{eqnarray}
with
\begin{eqnarray}\label{eq2}
T^{\text{tot}}_{ij} =T_{ij}+\beta\,\theta_{ij},
\end{eqnarray}
In this work, we adopt relativistic units where $G = c = 1$. The Ricci tensor is denoted by $R_{ij}$, and its contraction yields the Ricci scalar $\mathcal{R}$. The parameter $\beta$ appears in the field equations as a decoupling constant. The EMT is given by $T_{ij}$, while the additional source term $\theta_{ij}$ may represent contributions from other fields---such as scalar, vector, or tensor fields. To describe the interior spacetime of the stellar object, we consider a static, spherically symmetric line element:
\begin{eqnarray} \label{eq4}
 ds^2 = -e^ {\mathcal{S} (r)} dr^2 - r^2\big(d\theta^2 + \sin^2 \theta~ d\phi^2\big)~+ e^{\Phi( r)} dt^2 ,
 \end{eqnarray}
where the metric functions $\Phi$ and $\mathcal{S}$ are the only ones that are dependent on radial distance. We consider the internal composition of the stellar structure, defined by the total EMT $T^{\text{tot}}_{ij}$ indicates an anisotropic fluid as 
\begin{small}
\begin{eqnarray}
&& T^{\text{tot}}_{ij} = \left({\rho^{\text{tot}}}+{P^{\text{tot}}_t}\right)u_{i}u_{j}-{P^{\text{tot}}_t} g_{ij}+({P^{\text{tot}}_r}-{P^{\text{tot}}_t})\chi_{i}\,\chi_{j}~,\label{eq5}
\end{eqnarray}
\end{small}
here $u^i$ is found by solving $u^i=e^{\Phi(r)/2}{\delta^i}_4$, which gives us the four-velocity. In the radial direction, the unit vector is shown by the symbol $\chi^i$, whose value is equal to $e^{\mathcal{S}(r)/2} {\delta^i}_1$.
The radial and tangential pressures are denoted by $P^{\text{tot}}_r$ and $P^{\text{tot}}_t$, respectively, while the energy density is given by $\rho^{\text{tot}}$. The four-velocity vector $\chi^{i}$ satisfies the normalization condition $\chi^{i} \chi_{i} = -1$. Accordingly, the components of the total energy-momentum tensor $T^{\text{tot}}_{ij}$ can be expressed as:
\begin{eqnarray}\label{eq6}
[T^0_0]^{\text{tot}}=\rho^{\text{tot}}, ~~[T^1_1]^{\text{tot}}=-P^{\text{tot}}_{r},~~\text{and}~~[T^2_2]^{\text{tot}}=-P^{\text{tot}}_t.~~~
\end{eqnarray}
In this context, when $P^{\text{tot}}_r \neq P^{\text{tot}}_t$, the fluid distribution is anisotropic. The special case $P^{\text{tot}}_r = P^{\text{tot}}_t$ corresponds to an isotropic configuration. The anisotropy factor, denoted by $\Delta^{\text{tot}}$, is defined as $\Delta^{\text{tot}} = P^{\text{tot}}_t - P^{\text{tot}}_r$. The term $\frac{2(P^{\text{tot}}_t - P^{\text{tot}}_r)}{r}$ represents the force due to anisotropy within the fluid. If $P^{\text{tot}}_t > P^{\text{tot}}_r$, this force acts outward; if $P^{\text{tot}}_t < P^{\text{tot}}_r$, it acts inward. An outward-directed anisotropic force, where $P^{\text{tot}}_t$ exceeds $P^{\text{tot}}_r$, can support a more compact stellar configuration compared to the isotropic case~\cite{Gokhroo1994}.

On the other hand, the mass function $m(r)$ for an anisotropic fluid configuration can be determined using the metric potential as
\begin{equation}
e^{-\mathcal{S}} = 1 - \frac{2m(r)}{r}.\label{eq12}
\end{equation}
Additionally, one may get another version of mass formula~\cite{Sharp-Misner1964} 
\begin{equation}
m(r)= \frac{8\pi}{2} \int \rho^{\text{tot}}\, r^2 dr.\label{eq14}
\end{equation}
{Equations (\ref{eq12}) and (\ref{eq14}) are used to get}
\begin{equation}
\frac{\Phi^\prime }{2}= \frac{8\pi r \,P^{\text{tot}}_r + 2m/r^2}{1 - 2m/r }.\label{eq15}
\end{equation}
Now let us write the comprehensive representation of Einstein's field equations may be interpreted as the set of the subsequent differential equations associated with the metric~(\ref{eq1}): 
\begin{small}
\begin{eqnarray}
 \frac{1}{8\pi}\bigg[-\frac{1}{r^2}+e^{-\mathcal{S}}\left(\frac{1}{r^2}+\frac{\Phi^{\prime}}{r}\right)\bigg]&=& P^{\text{tot}}_r ,\label{eq9}\\
 \frac{1}{8\pi}\bigg[\frac{e^{-\mathcal{S}}}{4}\left(2\Phi^{\prime\prime}+\Phi^{\prime2}-\mathcal{S}^{\prime}\Phi^{\prime}+2\frac{\Phi^{\prime}-\mathcal{S}^{\prime}}{r}\right)\bigg]&=&  P^{\text{tot}}_t ,~~\label{eq10} \\
 \frac{1}{8\pi}\bigg[\frac{1}{r^2}-e^{-\mathcal{S}}\left(\frac{1}{r^2}-\frac{\mathcal{S}^{\prime}}{r}\right)\bigg]&=&  \rho^{\text{tot}}.\label{eq11}
\end{eqnarray}
\end{small}
The conservation of the total EMT ($T^{\text{tot}}_{ij}$ ) is required by the Einstein tensor $(G_{ij})$ in order to satisfy the Bianchi identity i.e.  i.e. $ {\nabla_i}[\,T^{ij}\,]^{\text{tot}}=0$ that gives pressure gradient, 
\begin{small}
\begin{equation}
\frac{dP^{\text{tot}}_r}{dr} =- \frac{8\pi r P^{\text{tot}}_r + 2m/r^2}{1 -2m/r } (P^{\text{tot}}_r+\rho^{\text{tot}})+ \frac{2 (P^{\text{tot}}_t-P^{\text{tot}}_r)}{r}.\label{eq7}
\end{equation}
\end{small}
This leads to the modified hydrostatic TOV equation for anisotropic stellar configurations~\cite{TOV1, TOV2}.

Our primary objective is to find an exact solution to the field equations (\ref{eq9})--(\ref{eq11}) associated with the TOV equation (\ref{eq7}), which governs the structure of a SS. Due to the highly nonlinear nature of these equations, obtaining a direct solution is quite challenging. To address this, we adopt an alternative method known as gravitational decoupling via the minimal geometric deformation (MGD) approach, which relies on a specific transformation of the gravitational potential,
\begin{eqnarray}
&& \Phi(r) \longrightarrow \mathcal{A}(r)+\beta\, h(r), \label{eq26}\\
&& e^{-\mathcal{S}(r)} \longrightarrow \mathcal{B}(r)+\beta\, g(r).  \label{eq27}
\end{eqnarray}

Let $h(r)$ and $g(r)$ represent the deformation functions corresponding to the temporal and radial components of the metric, respectively. The extent of this deformation can be controlled by adjusting the decoupling constant $\beta$. When $\beta = 0$, the framework smoothly reduces to standard GR. In the context of the MGD approach, it is found that $h(r) = 0$ and $g(r) \neq 0$, indicating that the deformation affects only the radial part of the metric, while the temporal component remains unchanged. This technique effectively splits the full system of equations (\ref{eq9})--(\ref{eq11}) into two independent sectors: the primary one associated with the EMT $T_{ij}$, and the secondary one linked to the additional source $\theta_{ij}$. To construct the initial system, we consider the energy-momentum tensor $T_{ij}$ describing an anisotropic matter distribution, given by:
\begin{equation}\label{eq28}
T_{ij}=\left(\rho+P_t\right)u_{i}\,u_{j}-P_t\,g_{ij}+\left(P_r-P_t\right)\chi_{i}\,\chi_{j}.
\end{equation}
The energy density is represented by $\rho$, while $P_r$ and $P_t$ denote the radial and tangential pressures of the seed solution, respectively. Based on these, the effective physical quantities can be written as
\begin{eqnarray}
\rho^{\text{tot}}=\rho+\beta \,\theta^0_0,~~P^{\text{tot}}_r=P_r-\beta\,\theta^1_1,~~P^{\text{tot}}_t=P_t-\beta\,\theta^2_2.~~~~  \label{eq29}
\end{eqnarray}

Furthermore, the related total anisotropy is
\begin{eqnarray} 
&&\hspace{-0.7cm} \Delta^{\text{tot}}=P^{\text{tot}}_t-P^{\text{tot}}_r= \Delta_{GR}+\Delta_{\theta},
 \label{eq30}\\
&&\hspace{-0.7cm} \text{where}~~~~\Delta_{GR}= P_t-P_r~~~~~\text{and}~~~~~\Delta_\theta= \beta (\theta^1_1-\theta^2_2)\nonumber.
\end{eqnarray}
The total anisotropy is defined as the combined effect of the anisotropies associated with the sources $T_{ij}$ and $\theta_{ij}$. The additional anisotropy, denoted by $\Delta_{\theta}$, arises due to gravitational decoupling and can enhance the original (unperturbed) anisotropy. Applying the transformations (\ref{eq26}) and (\ref{eq27}), the system of equations (\ref{eq9})--(\ref{eq11}) can be separated into two distinct subsystems. When $\beta = 0$, the first subsystem depends solely on the gravitational potentials $\mathcal{A}$ and $\mathcal{B}$:
\begin{eqnarray}
&&\hspace{-0.8cm}\rho= \frac{1}{8\pi} \bigg(\frac{1 }{r^2}-\frac{\mathcal{B}  }{r^2}-\frac{\mathcal{B}^{\prime} }{r} \bigg),\label{eq19}\\
&&\hspace{-0.8cm} P_r=\frac{1}{8\pi} \bigg(-\frac{1 }{r^2}+\frac{\mathcal{B}  }{r^2}+\frac{\mathcal{A}^{\prime} \mathcal{B}  }{r}\bigg), \label{eq2.20}\\
&&\hspace{-0.8cm} P_t=\frac{1}{8\pi} \bigg(\frac{\mathcal{B}^{\prime} \mathcal{A}^{\prime}  }{4}+\frac{\mathcal{A}^{\prime \prime} \mathcal{B}  }{2}+\frac{\mathcal{A}^{\prime 2} \mathcal{B} }{4}  +\frac{\mathcal{B}^{\prime}  }{2 r}+\frac{\mathcal{A}^{\prime} \mathcal{B}  }{2 r}\bigg). \label{eq21}
\end{eqnarray}

Using Equation (\ref{eq7}), a specific result is derived: 
\begin{eqnarray}
-P_r^{\prime}-\frac{\mathcal{A}^\prime}{2}(\rho+P_r)-\frac{2}{r}( P_{r}-P_t)=0.~~\label{eq22}
\end{eqnarray}

A solution to the TOV equation (\ref{eq19})-(\ref{eq21}) for the structure of the system can potentially be determined in the spacetime that is shown as follows: 
\begin{equation}\label{eq35}
ds^2=e^{\mathcal{A}(r)}dt^2-\frac{dr^2}{\mathcal{B}(r)}-(r^2d\theta^2+r^2\text{sin}^2\theta \,d\phi^2).
\end{equation}

One may get a second set of mathematical equations by turning on $\beta$ as 
\begin{small}
\begin{eqnarray}
&&\hspace{-0.7cm}\theta^{0}_0=-\frac{1}{8\pi}\Big(\frac{g   }{r^2}+\frac{g^\prime }{r}\Big), \label{eq36}\\
&&\hspace{-0.7cm}\theta^1_1=-\frac{1}{8\pi}\Big(\frac{g  }{r^2}+\frac{\mathcal{A}^{\prime} g   }{r}\Big), \label{eq37}\\
&&\hspace{-0.7cm}\theta^2_2=- \frac{1}{8\pi}\Big(\frac{1}{4} g^\prime \mathcal{A}^{\prime}   +\frac{1}{2} \mathcal{A}^{\prime \prime} g   +\frac{1}{4} \mathcal{A}^{\prime 2} g  +\frac{g^\prime  }{2 r}+\frac{\mathcal{A}^{\prime} g  }{2 r}\Big). \label{eq38}
\end{eqnarray}
\end{small}
The linear combination of the equations (\ref{eq36})-(\ref{eq38}) yields the subsequent resultant equation as 
\begin{eqnarray}
-\frac{\mathcal{A}^{\prime}}{2} (\theta^0_0-\theta^1_1)+ (\theta^1_1)^\prime+\frac{2}{r} ~(\theta^1_1-\theta^2_2)=0. \label{eq39}
\end{eqnarray}

Using this method, we can estimate the mass distribution across every system, 
\begin{small}
\begin{eqnarray}
&&\hspace{-0.7cm} m_{Q}=\frac{8\pi}{2} \int^r_0 \rho(x)\, x^2 dx~~~\text{and}~~~m_{\theta}= \frac{8\pi}{2}\,\int_0^r \theta^0_0 (x)\, x^2 dx~. \label{eq40}
\end{eqnarray}
\end{small}
The mass functions corresponding to the sources $T_{ij}$ and $\theta_{ij}$ are denoted by $m_{GR}(r)$ and $m_{\theta}(r)$, respectively. The strength of the MGD-decoupling approach lies in its ability to extend any known solution for the matter sector $\{T_{ij}, \mathcal{A}, \mathcal{B}\}$, as described by Eqs.~(\ref{eq9})--(\ref{eq11}), while also enabling the treatment of the more complex gravitational structure defined by Eqs.~(\ref{eq36})--(\ref{eq38}) to determine $\{\theta_{ij}, h, g\}$. As a result, it becomes possible to construct the corresponding "$\theta$-version" of any existing $\{T_{ij}, \mathcal{A}, \mathcal{B}\}$ solution as
\begin{eqnarray}
\{T_{ ij},~ \mathcal{A}(r),~ \mathcal{B}(r)\} \Longrightarrow \{T^{\text{tot}}_{ ij}, ~\Phi(r),~~\mathcal{S}(r)\}.
\end{eqnarray}
This connection defines a straightforward method to examine the effects of gravity that beyond conventional Einsteinian gravity. The static metric tensor in GR will be modified as below:\\
\begin{equation}
g_{ij} \Longrightarrow g_{ij}^{\text{tot}} = \begin{bmatrix}
e^{\mathcal{A}} & 0 & 0 & 0\\
0 & {1 \over \mathcal{B}+\beta\,g} & 0 & 0\\
0 & 0 & r^2 & 0\\
0 & 0 & 0 & r^2 \sin^2\theta
\end{bmatrix}
\end{equation}
which can be seen at very small $\beta$ (or ~$\beta g/\mathcal{B}<<1)$ as
\begin{footnotesize}
\begin{eqnarray}
\hspace{-0.5cm} g_{ij}^{\text{tot}}= g_{ij}+\beta \,h_{ij} \approx  
\begin{bmatrix}
e^{\mathcal{A}} & 0 & 0 & 0\\
0 & {1 \over \mathcal{B}} & 0 & 0\\
0 & 0 & r^2 & 0\\
0 & 0 & 0 & r^2 \sin^2\theta
\end{bmatrix}
-\beta \begin{bmatrix}
0 & 0 & 0 & 0\\
0 & {g \over \mathcal{B}^2} & 0 & 0\\
0 & 0 & 0 & 0\\
0 & 0 & 0 & 0
\end{bmatrix}~.
\end{eqnarray}
\end{footnotesize}
Hence, the functions in the metric elements i.e. $[\mathcal{A}(r),\, \mathcal{B}(r),\, g(r)]$ are scalar functions of $r$ only.

\section{Perturbed SS model through MGD } \label{sec3}
For modeling the SS, we use the MIT bag model EOS \cite{Chodos:1974}, which describes the distribution of strange quark matter (SQM) inside the star. The inclusion of the bag constant in this model accounts for all modifications to the energy density and pressure of the SQM. Our simplified bag model assumes that the quarks are massless and non-interacting. Under these assumptions, the quark pressure is given by
\begin{eqnarray} \label{eq41}
P_r=\sum_{f=u,~d,~s} P^{f}-\mathcal{B}_g.
\end{eqnarray}

Let $P^f$ represent the pressures of the individual quark flavors $u$, $d$, and $s$. The total external pressure, known as the Bag constant $\mathcal{B}_g$, acts to balance these quark pressures. Within the framework of the MIT Bag model, the energy density $\rho$ of the deconfined quarks can be expressed as:
\begin{eqnarray}
\rho=\sum_{f}\rho^{f}+\mathcal{B}_g,~~~\text{where}~~~\rho^f=3P^f.\label{eq42}
\end{eqnarray}
Using Eqs.~(\ref{eq41}) and (\ref{eq42}), together with the relation $\rho^f = 3P^f$, we can express the MIT bag EOS for strange quark stars in its exact form as 
\begin{eqnarray} \label{eq2.4}
P_r=\frac{1}{3}(\rho-4\mathcal{B}_g).
\end{eqnarray} 
At this juncture, it is essential to identify the realistic matter density to determine the pressure for the bag model. In this context, we examine a non-singular, monotonically declining matter density inside the spherically symmetric stellar system described by Mak and Harko~\cite{Harko:2002pxr} as:
 \begin{eqnarray}
     \rho(r)=\rho_0 \bigg[ 1-\bigg(1-\frac{\rho_s}{\rho_0}\bigg) \frac{r^2}{r^2_s} \bigg]. \label{eq2.1}
 \end{eqnarray}
The central and surface values of $\rho$ are denoted by the variables $\rho_0$ and $\rho_s$, correspondingly.  Once the matter density from Eq. (\ref{eq2.1}) is entered into the MIT bag model EOS (\ref{eq2.4}), the bag pressure $P_r$ is calculated as:
\begin{eqnarray}
 &&\hspace{-0.5cm} P_r=\frac{r_s^2 (\rho_0-4 \mathcal{B}_g)+r^2 (\rho_s-\rho_0)}{3 r_s^2}. \label{eq3.5}
\end{eqnarray}
Now we shall focus on finding the spacetime geometries $\mathcal{A}$ and $\mathcal{B}$ for the unperturbed system (\ref{eq19})-(\ref{eq21}). There, a differential equation is derived by using Eqs. (\ref{eq2.1}) and (\ref{eq19}) as:  
\begin{eqnarray}
\hspace{-0.7cm} r \mathcal{B}'(r)+\mathcal{B}(r)+\frac{r^4 (\rho_s-\rho_0)+r^2 \rho_0 r_s^2-8 \pi  r_s^2}{8 \pi  r_s^2}=0~.
\end{eqnarray}
Upon completing the integration, we determine the potential $\mathcal{B}$ in the following manner:
\begin{eqnarray}
 \mathcal{B}=\frac{8 \pi  r^4 (\rho_0-\rho_s)}{5 r_s^2}-\frac{8}{3} \pi  \rho_0 r^2+1. \label{eq3.7a}
\end{eqnarray}
To find the other potential $\mathcal{A}$, we combine Eqs. (\ref{eq3.5}), (\ref{eq3.7a}) and (\ref{eq2.20}), which yields a subsequent differential equation
\begin{eqnarray}
&& 5 r_s^2 (32 \pi \mathcal{B}_g r+3 \mathcal{A}^\prime)+8 \pi \rho_0 r \big(3 \mathcal{A}^\prime r^3-5 \mathcal{A}^\prime r r_s^2+8 r^2 \nonumber \\
&& -10 r_s^2\big)-8 \pi \rho_s r^3 (3 \mathcal{A}^\prime r+8)=0, \label{eq3.8}
\end{eqnarray}
Integrating above Eq.~(\ref{eq3.8}) that yields a following solution for $\mathcal{A}(r)$ as
\begin{eqnarray}
\mathcal{A}(r) &=&\frac{2 \sqrt{10 \pi } r_s (\rho_0-6\mathcal{B}_g)}{3 \sqrt{10 \pi \rho_0^2 r_s^2-9\rho_0+9\rho_s}}~\tanh ^{-1}\left[\frac{\sqrt{\frac{2 \pi }{5}} \left(5\rho_0 r_s^2+6\rho_s r^2-6\rho_0 r^2\right)}{r_s \sqrt{10 \pi \rho_0^2 r_s^2-9\rho_0+9\rho_s}}\right]\nonumber \\
&&-\frac{2}{3} \ln \big[8 \pi \rho_0 r^2 \left(3 r^2-5 r_s^2\right)-24 \pi \rho_s r^4  +15 r_s^2\big]+\mathcal{F}=\Phi (r).  \label{eq3.9}
\end{eqnarray}
This set of equations (\ref{eq3.7}) and (\ref{eq3.9}) provides the complete spacetime geometry for the unperturbed solution for the system (\ref{eq19})-(\ref{eq21}), and the corresponding other pressure component is calculated as, 
\begin{eqnarray}
&&\hspace{-0.5cm} P_t= -\frac{1}{3 r_s^2 \left(-24 \pi  \rho_0 r^4+40 \pi  \rho_0 r^2 r_s^2+24 \pi  \rho_s r^4-15 r_s^2\right)} \Big[\rho_0 r_s^2 \Big(128 \pi  \mathcal{B}_g r^4-10 r^2  (8 \pi \mathcal{B}_g r_s^2+3)\nonumber\\
&&\hspace{0.2cm}+15 r_s^2\Big) +2 \rho_s r^2 r_s^2  \left(15-64 \pi  \mathcal{B}_g r^2\right)+20 \mathcal{B}_g r_s^4 \left(8 \pi  \mathcal{B}_g r^2-3\right) +8 \pi  \rho_0^2 r^2(2 r^4-5 r^2 r_s^2+5 r_s^4)\nonumber\\
&&\hspace{0.2cm}-8 \pi  \rho_0 \rho_s r^4 \left(4 r^2-5 r_s^2\right)+16 \pi  \rho_s^2 r^6\Big]
\end{eqnarray}
while the unperturbed spacetime can be given as,
\begin{eqnarray}
    \label{eq35a}
&& \hspace{-0.3cm} ds^2=\Bigg[\frac{2 \sqrt{10 \pi } \,r_s (\rho_0-6\mathcal{B}_g)}{3 \sqrt{10 \pi \rho_0^2 r_s^2-9\rho_0+9\rho_s}}~\tanh^{-1} \left(\frac{\sqrt{\frac{2 \pi }{5}} \left(5\rho_0 r_s^2+6\rho_s r^2-6\rho_0 r^2\right)}{r_s \sqrt{10 \pi \rho_0^2 r_s^2-9\rho_0+9\rho_s}}\right)\nonumber\\
&&\hspace{0.5cm}-\frac{2}{3} \ln \big[8 \pi \rho_0 r^2 \left(3 r^2-5 r_s^2\right)-24 \pi \rho_s r^4+15 r_s^2\big]+\mathcal{F}\Bigg]dt^2\nonumber\\
&&\hspace{0.5cm}-\Bigg[\frac{8 \pi  r^4 (\rho_0-\rho_s)}{5 r_s^2}-\frac{8}{3} \pi  \rho_0 r^2+1\Bigg]^{-1}dr^2-r^2 d\Omega^2.~~~~~
\end{eqnarray}
where, $d\Omega^2=d\theta^2+\text{sin}^2\theta \,d\phi^2$. However, to fully understand the $\theta$-sector, we need to solve the second set of equations (\ref{eq36})--(\ref{eq38}). It is important to note that this system consists of three independent equations but involves four unknown variables.
Since we are interested in finding the solution to the second system under a perturbation approach. Imagining a perturbation due to tical deformation from a neighboring object then the metric is perturb as $g_{\mu \nu}^{tot} = g_{\mu \nu}+h_{\mu \nu}$, where $g_{\mu \nu}$ is the unperturbed spacetime metric and $h_{\mu \nu}$ is the perturbation. In general, perturbation due to tidal deformation in linearized static scenario can be written as \citep{reg,thor}
\begin{small}
\begin{eqnarray}
&& h_{\mu \nu} = \text{diag}\big[\big\{e^\nu H(r),-e^\lambda H(r),-r^2 H(r),-r^2 H(r)\sin^2\theta \big\} \nonumber \\
&& \hspace{2cm} Y_{2m}(\theta,\phi)\big]~.
\end{eqnarray}
\end{small}
The above choice satisfy the Regge-Wheeler gauge \citep{reg} with $l=2$ i.e. even-parity perturbation. The perturbation leads to
$\delta T^0_0 =\delta \rho=(dP_r/d\rho)^{-1}\delta P_r$ and $\delta G^\theta_\theta+\delta G^\phi_\phi=16\pi \,\delta P_r$, which implies to a differential equation in $H(r)$. As a consequence, we have chosen the perturbation as a function of radial coordinate $r$ only \citep{hin}. Consequently, let us consider a perturbation caused by an external force (such as gravitational waves or accretion) that minimally distorts the interior spacetime in the manner of a harmonic oscillation, represented as
\begin{eqnarray}
   g(r)= \sin \,(\Psi r^2) 
\end{eqnarray}
where the parameter $\Psi$ is considered as the frequency of the perturbation along the radial direction. Physically, $\Psi$ does not represent a temporal frequency or amplitude; rather, it  sets the radial scale of the perturbation inside the star. In a compact star, oscillations arise from  the balance between gravitational pull and pressure gradients. The quantity $\Psi$ determines how quickly these perturbations vary with radius: a larger $\Psi$ corresponds to more rapid spatial oscillations, while a smaller $\Psi$ gives more gradual changes. Essentially, $\Psi$ encodes the ``stiffness''  of the stellar medium to radial perturbations--how strongly the restoring forces react to deviations from equilibrium. This makes $\Psi$ a measure of the radial wavelength of the perturbation,  reflecting the internal structure and density profile of the star, rather than a conventional frequency or amplitude. The relation between stellar mass and tidal deformability $(M-\Lambda)$ comes from how a star's matter reacts to an external tidal field via the perturbed spacetime metric. The tidal deformability \cite{hin, Hinderer:2009ca} is expressed as
\begin{equation}
\Lambda = \frac{2}{3} \,k_2(C, y_{_R}) \,C^{-5}, \quad y_{_R} = \frac{R\, H'(R)}{H(R)}, \quad C = \frac{M}{R}~, 
\end{equation}
where $k_2$ is the quadrupolar Love number, $y_{_R}$ measures the slope of the radial perturbation at the surface, and $C$ is the compactness. In the usual static even-parity $l=2$ framework, the radial function $H(r)$ satisfies a second-order differential equation whose coefficients depend on the star's internal density and pressures. This sets up the chain
\begin{equation}
h_{\mu\nu}(H(r)) \;\rightarrow\; \delta G_{\mu\nu} \;\rightarrow\; \delta T_{\mu\nu} \;\rightarrow\; \{\delta \rho, \delta P_r, \delta P_t\}~, 
\end{equation}
linking the external tidal field to the star's matter and, ultimately, to $\Lambda$. To make this tractable analytically while keeping the key physics, we use the radial harmonic form
\begin{equation}
H(r) \approx g(r) = \sin(\Psi r^2)~, 
\end{equation}
which represents a standing wave pattern of tidal response inside the star. Its derivatives are
\begin{small}
\begin{equation}
g'(r) = 2 \Psi r \cos(\Psi r^2), ~~ g''(r) = 2 \Psi \cos(\Psi r^2) - 4 \Psi^2 r^2 \sin(\Psi r^2), 
\end{equation}
\end{small}
showing that $g(r)$ satisfies an oscillator-like radial equation
\begin{equation}
g''(r) + 4 \Psi^2 r^2 g(r) \approx 2 \Psi \cos(\Psi r^2)~, 
\end{equation}
and expands regularly at the center as
\begin{equation}
g(r) \sim \Psi r^2 - \frac{(\Psi r^2)^3}{3!} + \dots~~, 
\end{equation}
highlighting nodes typical of radial standing waves. The angular part is handled by the spherical harmonic
\begin{equation}
\Delta_\Omega Y_{2m}(\theta, \phi) = -6 Y_{2m}(\theta, \phi)~, 
\end{equation}
so that the full perturbation is written as
\begin{equation}
h_{\mu\nu}(r, \theta, \phi) \sim g(r) Y_{2m}(\theta, \phi)~, 
\end{equation}
where the angular harmonic fixes the quadrupolar shape and the radial harmonic controls how strongly the deformation penetrates different layers. At the surface, the logarithmic derivative is
\begin{equation}
y_{_R} = \frac{r\, g'(r)}{g(r)}\Big|_{r=R} = \frac{2 \Psi R^2 \cos(\Psi R^2)}{\sin(\Psi R^2)}~, 
\end{equation}
and the mass is determined by
\begin{equation}
M = 4 \pi \int_0^R \rho(r)\, r^2 dr~. 
\end{equation}
Putting it together, the tidal deformability as a function of mass and the radial harmonic parameter is
\begin{equation}
\Lambda(M, \Psi) = \frac{2}{3} \,C^{-5} k_2 \left(C, \frac{2 \Psi R^2 \cos(\Psi R^2)}{\sin(\Psi R^2)}\right)~, 
\end{equation}
demonstrating that the $M-\Lambda$ relation arises from the tidal field deforming the curvature, the radial harmonic $g(r)$ transmitting this effect through the star's matter, and the surface phase of the radial mode setting the Love number. Choosing $g(r) = \sin(\Psi r^2)$ is therefore not arbitrary; it is a controlled analytic approximation of the exact radial solution, maintaining regularity at the center, proper separation between radial and angular harmonics, and the correct link between matter and geometry that determines the observable $M-\Lambda$ behavior. In our model, $g(r) = \sin(\Psi r^2)$ is chosen as {\it harmonic} function because it represent oscillations caused by the interplay between gravity and pressure inside the star. These forces act like a restoring mechanism, producing wave-like behavior similar to harmonic motion in familiar physical systems. Even though the sine depends on $r^2$, so the ``frequency'' changes with radius, the oscillations are still sinusoidal. Physically, this reflects that deeper layers of the star experience stronger gravitational and pressure forces, which naturally change the oscillation pattern. So, calling it harmonic refers to the underlying physical mechanism generating these oscillations, rather than implying a strictly constant-frequency wave. Then the component of $\theta$-sector are,
\begin{eqnarray}
&&\hspace{-0.cm} \theta^0_0=-\frac{\sin \left(r^2 \Psi \right)+2 r^2 \Psi  \cos \left(r^2 \Psi \right)}{8 \pi  r^2},\\
&&\hspace{-0.cm} \theta^1_1=\frac{5 \sin \left(r^2 \Psi \right)}{8 \pi  r^2 \left(8 \pi \rho_0 r^2 \left(3 r^2-5 r_s^2\right)-24 \pi \rho_s r^4+15 r_s^2\right)} \Big[r_s^2 \Big(32 \pi \mathcal{B}_g r^2-3\Big)
\nonumber\\&&\hspace{0.5cm}+8 \pi \rho_0 r^2 \left(r^2-r_s^2\right) -8 \pi \rho_s r^4\Big],\\
&&\hspace{-0.cm} \theta^2_2=\frac{}{8 \pi  \left(8 \pi \rho_0 r^2 \left(3 r^2-5 r_s^2\right)-24 \pi \rho_s r^4+15 r_s^2\right)^2}\Big[\Psi  \cos \left(r^2 \Psi \right) \big(8 \pi \rho_0 r^2 \left(3 r^2-5 r_s^2\right)\nonumber\\&&\hspace{0.5cm} -24 \pi \rho_s r^4+15 r_s^2\big) \big\{5 r_s^2    \left(16 \pi \mathcal{B}_g r^2-3\right)+8 \pi \rho_0 r^4-8 \pi \rho_s r^4\big\}-16 \pi  \sin \left(r^2 \Psi \right) \nonumber\\
&&\hspace{0.5cm}\Big\{5\rho_0 r_s^2 \big(112 \pi \mathcal{B}_g r^4-8 r^2   \left(10 \pi \mathcal{B}_g r_s^2+3\right) +15 r_s^2\big) +40\rho_s r^2 r_s^2 \left(3-14 \pi \mathcal{B}_g r^2\right) \nonumber\\
&&\hspace{0.5cm} +50 \mathcal{B}_g r_s^4 \left(8 \pi \mathcal{B}_g r^2-3\right)+4 \pi \rho_0^2 r^2 \left(16 r^4-30 r^2 r_s^2+25 r_s^4\right) \nonumber\\
&&\hspace{0.5cm} -8 \pi \rho_0\rho_s r^4 (16 r^2-15 r_s^2)+64 \pi \rho_s^2 r^6\Big\}\Big]
\end{eqnarray}
and corresponding perturbed or deformed spacetime can be described by the following line element,
\begin{eqnarray}
&& ds^2=\Bigg[\frac{2 \sqrt{10 \pi } \,r_s (\rho_0-6\mathcal{B}_g) }{3 \sqrt{10 \pi \rho_0^2 r_s^2-9\rho_0+9\rho_s}}~\tanh ^{-1}\left(\frac{\sqrt{\frac{2 \pi }{5}} \left(5\rho_0 r_s^2+6\rho_s r^2-6\rho_0 r^2\right)}{r_s \sqrt{10 \pi \rho_0^2 r_s^2-9\rho_0+9\rho_s}}\right)\nonumber\\&&\hspace{0.5cm}-\frac{2}{3} \ln \big\{8 \pi \rho_0 r^2 \left(3 r^2-5 r_s^2\right) -24 \pi \rho_s r^4 +15r_s^2\big\}+ \mathcal{F}\Bigg]dt^2 \nonumber\\&&\hspace{0.5cm}-\Bigg[\frac{8 \pi  r^4 (\rho_0-\rho_s)}{5 r_s^2}-\frac{8}{3} \pi  \rho_0 r^2+1+\beta \sin \,(\Psi r^2) \Bigg]^{-1}dr^2-d\Omega^2.
\end{eqnarray}
The next sections shall focus on discussing the boundary conditions and physical analysis of the perturbed SS solution. 

\section{Boundary conditions for perturbed decoupled SS models}\label{sec4} 

To model a realistic dense celestial object with a finite matter distribution characterized by a specific mass $M$ and radius $R$, it is essential to match the interior geometry $\mathcal{M}^-$ at the boundary $r = R$ with the exterior spacetime $\mathcal{M}^+$. In the context of GR, the exterior is typically described by the Schwarzschild vacuum solution, especially for uncharged, non-radiating, and static compact stars. However, when considering the additional $\theta$-sector, its influence on the exterior spacetime must be taken into account. As a result, the exterior metric can be expressed by the following line element:
\begin{eqnarray}
ds^{2} &=& \left[1-\frac{2{\mathcal{M}}}{r}\right]dt^{2}- \bigg[1-\frac{2{\mathcal{M}}}{r}+\beta\, \mathcal{D^\ast}(r)\bigg]^{-1}dr^2 \nonumber \\
&& -r^{2}d\Omega^{2}. \label{eq3.1}
\end{eqnarray}
The function $g^\ast(r)$ represents the deformation of the exterior Schwarzschild spacetime caused by the additional source $\theta_{ij}$. The metric in Eq.~(\ref{eq3.1}) describes a Schwarzschild spacetime that is deformed and thus no longer vacuum. However, it is possible to set $g^\ast(r)$ to zero to recover the standard exterior vacuum solution, simplifying the model without losing generality.  To properly match the interior and exterior regions, the junction conditions require satisfying both the first and second fundamental forms. The first fundamental form ensures the continuity of the metric functions across the boundary at $r = r_s$. This condition can be expressed as:
\begin{equation}\label{eq3.2}
e^{\mathcal{S}^{-}(r)}|_{r=r_s}=e^{\mathcal{S}^{+}(r)}|_{r=r_s},
\end{equation}
and
\begin{equation}\label{eq3.3}
e^{\mathcal{A}^{-}(r)}|_{r=r_s}=e^{\mathcal{A}^{+}(r)}|_{r=r_s},
\end{equation}
where $``-"$ denotes the inner geometry and $``+"$ denotes the exterior geometry. The second fundamental form refers the vanishing of total pressure at the boundary surface $r=r_s$ i.e., 
\begin{equation}\label{eq3.4}
\left[P_r^{(\text{tot})}(r)\right]_{r_s}=\left[P_r(r)-\beta\, \theta^{1}_{1}(r)\right]_{r_s}=0~,
\end{equation} 
that provides 
\begin{equation}
{P_r}(r_s)-\beta\,(\theta^1_1)^{-}(r_s)=-\beta\,(\theta^1_1)^{+}(r_s)~.~~~\label{eq3.5a}
\end{equation}
We are able to derive the resultant equation by looking at Eqs. (\ref{eq37}) and (\ref{eq3.5a}) as 
\begin{eqnarray}
{P}_r(r_s)+\frac{\beta}{8\pi}\,\bigg[g(r_s)\left(\frac{\mathcal{A}^{\prime}(r_s)}{r_s}+\frac{1}{r_s^{2}}\right)\bigg] =-\beta\,(\theta^1_1)^{+}(r_s).~~~~~~\label{eq3.6}
\end{eqnarray}
Now we need to find an expression for $(\theta^1_1)^{+}(r_s)$.  Since the component $(\theta^1_1)^{+}(r_s)$ corresponds to the outer spacetime, its value must be determined from the Schwarzschild exterior geometry. To obtain the expression for $(\theta^1_1)^{+}(r_s)$, we assume that $g^\ast$ represents the deformation function in the exterior spacetime. Then, using Eq.~(\ref{eq37}), we can write
\begin{eqnarray}
(\theta^1_1)^{+}(r_s)=-\frac{1}{8\pi}\Big(\frac{g^{\ast}(r_s)}{r_s^2}+\frac{(\mathcal{A}^{\prime})^+(r_s) g^{\ast} (r_s)   }{r_s}\Big)~, ~~~\label{eq52a}
\end{eqnarray}
From the line element (\ref{eq3.1}), we get 
\begin{eqnarray}
e^{\mathcal{A}^{+}(r_s)}= 1-\frac{2{\mathcal{M}}}{r_s}~, \label{eq53a}
\end{eqnarray}
Upon differentiating the above expression, we obtain 
\begin{eqnarray}
(\mathcal{A}^{\prime})^+(r_s)= \frac{2{\mathcal{M}}}{r^2_s} \left[1-\frac{2{\mathcal{M}}}{r_s}\right]^{-1}, \label{eq54a}
\end{eqnarray}
On substituting Eq. (\ref{eq54a}) in Eq. (\ref{eq53a}), we arrive at
\begin{eqnarray}
(\theta^1_1)^{+}(r_s)=-\frac{ g^{\ast}(r_s)}{8\pi}\Big(\frac{1}{r_s^2}+\frac{2{\mathcal{M}}}{r^3_s}\left[1-\frac{2{\mathcal{M}}}{r_s}\right]^{-1}\Big)~,~~~ \label{eq55a}
\end{eqnarray}
Finally, plugging the Eq.~(\ref{eq55a}), we find the final form of boundary condition,
\begin{eqnarray}
 && P_r(r_s)+\frac{\beta}{8\pi}\,\bigg[g(r_s)\left(\frac{\mathcal{A}^{\prime}(r_s)}{r_s}+\frac{1}{r_s^{2}}\right)\bigg] =\frac{\beta}{8\pi}~g^{\ast}(r_s) \nonumber \\
 && \hspace{3.5cm} \Bigg[\frac{2\mathcal{M}}{r^3_s\,\Big(1-2\mathcal{M}/r_s\Big)}+\frac{1}{r_s^2}\Bigg].~~~\label{eq3.7}
\end{eqnarray} 
The three main equations (\ref{eq3.2}), (\ref{eq3.3}) and (\ref{eq3.7}) describe the boundary conditions for establishing a minimally deformed perturbed solution for compact stars. For simplicity, if we assume that outer spacetime has no deformation, i.e. it is equivalent to the exact Schwarzschild metric,  then $g^{\ast}(r)$ must be zero in Eq.~(\ref{eq3.1}). In this case, the final form of the Eq.~(\ref{eq3.7}) gives, 
\begin{eqnarray}
\hspace{-0.5cm} P^{\text{tot}}_r(r_s)=P_r(r_s)+\frac{\beta}{8\pi}\,\bigg[g(r_s)\left\{\frac{\mathcal{A}^{\prime}(r_s)}{r_s}+\frac{1}{r_s^{2}}\right\}\bigg] =0.~~ \label{eq3.8a} 
 \end{eqnarray} 
 
Next, we will determine the arbitrary constant that is included in the solution by using the constraints (\ref{eq3.2}), (\ref{eq3.3}), and (\ref{eq3.8a}).  The following expressions for constants related to solution is provided in context of above conditions: 
\begin{eqnarray}
&&\hspace{-0.5cm}    \mathcal{B}_g=\frac{8 \pi  \rho_s r_s^2 \left(16 \pi  \rho_0 r_s^2+24 \pi  \rho_s r_s^2-15\right)-15 \beta  \left(8 \pi  \rho_s r_s^2+3\right) \sin \left(r_s^2 \Psi \right)}{32 \pi  r_s^2 \left(16 \pi  \rho_0 r_s^2+24 \pi  \rho_s r_s^2-15 \beta  \sin \left(r_s^2 \Psi \right)-15\right)},\\
&&\hspace{-0.5cm}    \mathcal{F}= \frac{2 \sqrt{10 \pi } \,r_s (\rho_0-6 \mathcal{B}_g) }{3 \sqrt{10 \pi  \rho_0^2 r_s^2-9 \rho_0+9 \rho_s}}~ \tanh ^{-1}\left(\frac{\sqrt{\frac{2 \pi }{5}} \,r_s (\rho_0-6 \rho_s)}{\sqrt{10 \pi  \rho_0^2 r_s^2-9 \rho_0+9 \rho_s}}\right)\nonumber\\&&\hspace{0.5cm}+\ln \bigg[1-\frac{16 \pi}{15}   \rho_0 r_s^2-  \frac{8}{5} \pi  \rho_s r_s^2 +\beta  \sin \left(r_s^2 \Psi \right)\bigg]\nonumber\\
&& \hspace{0.5cm} +\frac{2}{3} \ln \left\{15 r_s^2-8 \pi  r_s^4 (2 \rho_0+3 \rho_s)\right\}.
\end{eqnarray}

\section{Physical analysis of perturbed SS solution}\label{sec5}  
In this section, we investigate how thermodynamic variables behave inside compact stars, focusing on the effects of gravitational decoupling. Using a mix of analytical work and numerical simulations, we explore how the theory-driven parameters---namely, the deformation parameter $\beta$ and the perturbation parameter $\Psi$---affect the distribution of matter deep within these objects. These parameters play a key role in shaping how quantities like energy density and pressure evolve with radius. We would like to emphasize that the parameters $\beta$ and $\Psi$ are far from arbitrary; they are strictly constrained by both the MGD framework and the physical consistency of the stellar model. In this framework, the parameter $\beta$ controls the strength of the additional gravitational contribution introduced via the MGD of the radial metric. Setting $\beta = 0$ removes this deformation entirely, recovering the standard GR solution. Meanwhile, $\Psi$ sets the radial scale of the tidal perturbation, determining how quickly the deformation varies throughout the star. The allowed values for these parameters are determined by several physical requirements: the metric potentials must remain regular at the center; the energy density and pressures must be positive and decrease outward smoothly; all standard energy conditions must be satisfied; and the configuration must meet stability criteria, including causality and Herrera's cracking condition. These constraints ensure the model remains physically consistent while highlighting the boundaries of its applicability.

In particular, the causality principle provides a direct method to identify the upper bounds of $\beta$ and $\Psi$: as the deformation increases, the speed of sound inside the star rises, and the maximal allowable values correspond to the point where it reaches the speed of light. Exceeding these thresholds would result in unphysical behavior and signal a breakdown of the perturbative MGD approach. In summary, $\beta$ and $\Psi$ are not free parameters. $\beta$ quantifies the influence of the extra gravitational sector, while $\Psi$ sets the radial wavelength of the tidal perturbation. Their permissible ranges are determined by the chosen stellar model and EOS, ensuring that the star remains regular, stable, and physically realistic. We also examine how these perturbations influence broader stellar properties, particularly the $M-R$ relationship, which helps in comparing theoretical predictions with astrophysical data. Additionally, we consider how constraints on density introduced by the perturbations might affect the structure of the SSs, offering potential insights into the presence of exotic matter in their cores. To connect our theoretical framework with observational data, we examine some of the heaviest pulsars measured to date, which serve as critical benchmarks for testing the validity of stellar models under extreme conditions. Specifically, we consider the following high-mass NeStrs: PSR J0740+6620~\cite{NANOGrav:2019jur,Fonseca:2021wxt}, with $M = 2.08^{+0.07}_{-0.07},M_\odot$; PSR J1810+1744~\cite{PSRJ1810+1744}, at $M = 2.13^{+0.04}_{-0.04},M_\odot$; PSR J1959+2048~\cite{starPSR}, with $M = 2.18^{+0.09}_{-0.09},M_\odot$; and PSR J2215+5135~\cite{starPSR}, estimated at $M = 2.28^{+0.10}_{-0.09},M_\odot$. These compact objects provide strong constraints on the EOS of dense matter. For a stellar model to be physically credible, it must not only satisfy the governing equations but also align with observational evidence. This requires that the matter behave causally, the central density and pressure be finite and decrease smoothly toward the surface, and the configuration support at least the highest reliably measured NeStr masses. Failing any of these conditions would undermine the model's physical relevance.

\subsection{Energy density and pressure profiles behavior}

\begin{figure*}[!htp]
\centering
\includegraphics[height=6cm,width=7cm]{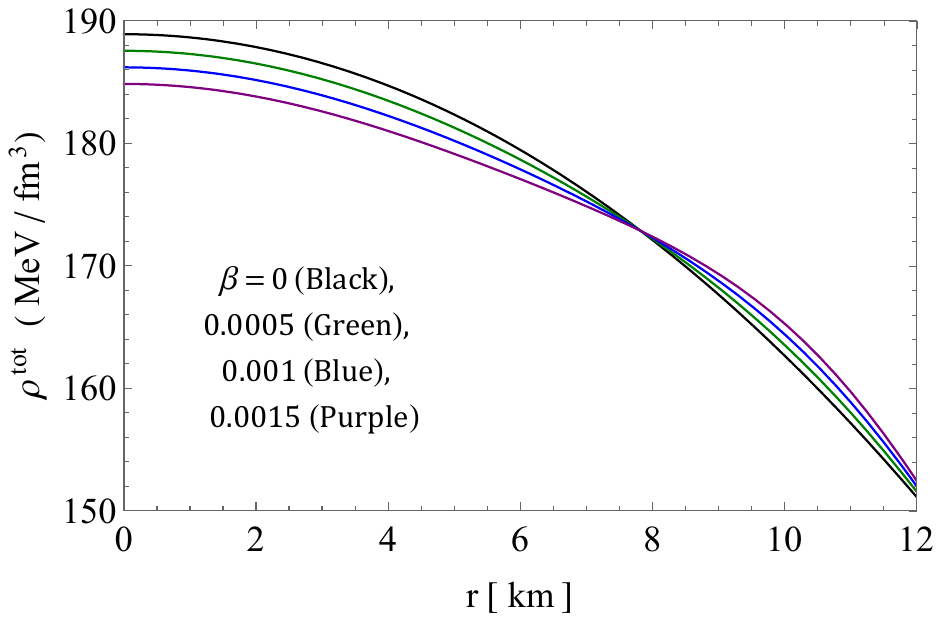}~~~~~
\includegraphics[height=6cm,width=7cm]{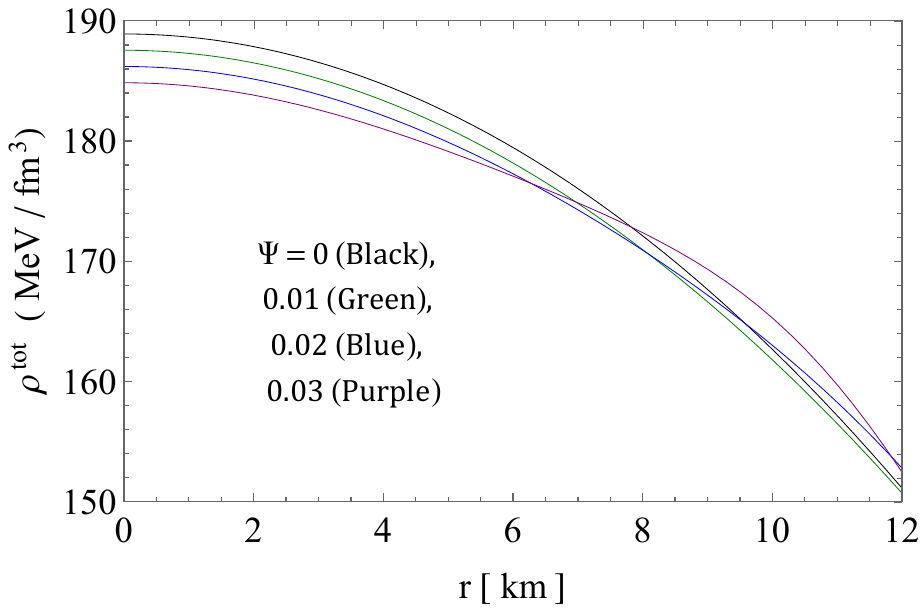}
\caption{Energy density ($\rho^{\text{tot}}$) against radial coordinate $r$ for different values of $\beta$ and $\Psi$ with $r_s =12\,km,\,\rho_0 =0.00025/km^2, \, \rho_s=0.00020/km^2,\, \Psi=0.03/km^2$ (left panel) and  $r_s =12\,km,\,\rho_0 =0.00025/km^2, \, \rho_s=0.00020/km^2,\, \beta=0.0015$ (right panel) respectively.}
\label{f1}
\end{figure*}

\begin{figure*}[!htp]
\centering
\includegraphics[height=6cm,width=7cm]{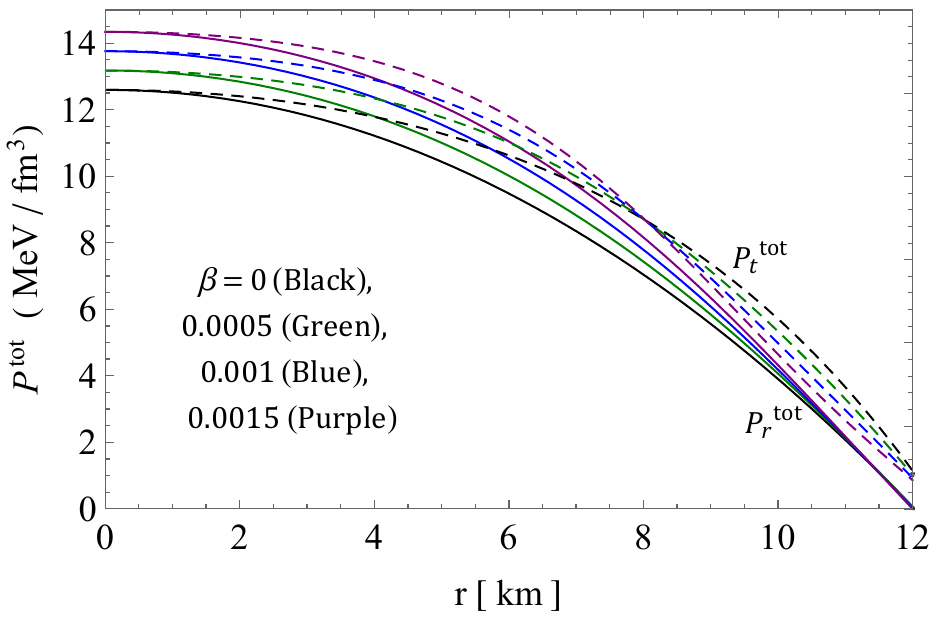}~~~~~
\includegraphics[height=6cm,width=7cm]{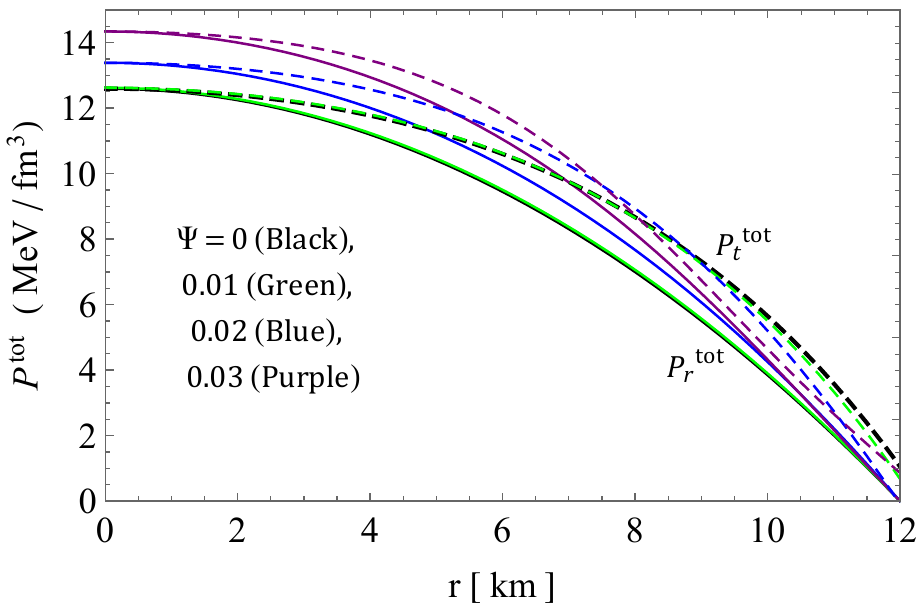}
\caption{Pressure ($P_r^{\text{tot}}$ \& $P_t^{\text{tot}}$) against radial coordinate $r$ for different values of $\beta$ and $\Psi$ with $r_s =12\,km,\,\rho_0 =0.00025/km^2, \, \rho_s=0.00020/km^2,\, \Psi=0.03/km^2$ (left panel) and  $r_s =12\,km,\,\rho_0 =0.00025/km^2, \, \rho_s=0.00020/km^2,\, \beta=0.0015$ (right panel) respectively.}
\label{f2}
\end{figure*}

For a compact star to be physically plausible, its internal matter must exhibit uniform behavior throughout the stellar interior. The energy density must consistently be positive throughout, ensuring the existence of real matter throughout the model. Likewise, both $P_r^{\text{tot}}$ and $P_t^{\text{tot}}$ must remain positive to counterbalance the gravitational force of the star. At the core, these quantities must remain finite and exhibit smooth variation, devoid of singularities or irregularities, ensuring a stable and physically consistent internal structure. In summary, $\rho^{\text{tot}}$, $P_r^{\text{tot}}$, and $P_t^{\text{tot}}$ must be positive and decrease gradually outward to sustain a plausible stellar configuration. Figures \ref{f1} and \ref{f2} illustrate that the energy density and pressures attain their peak values at the core and progressively decrease toward the surface. This trend is characteristic of a stable, self-gravitating compact star, wherein the core region undergoes the greatest compression. As one moves outward, the radialpressures consistently decrease and ultimately vanish at the stellar surface, satisfying the zero-pressure boundary condition necessary to equilibrate internal forces with the external vacuum.  This trend continues across many values of the deformation parameter $\beta$ and the perturbation parameter $\Psi$, suggesting that these modifications have no impact on the star's typical structural configuration.


\subsection{Anisotropic profile behavior}

\begin{figure*}[!htp]
\centering
\includegraphics[height=6cm,width=7cm]{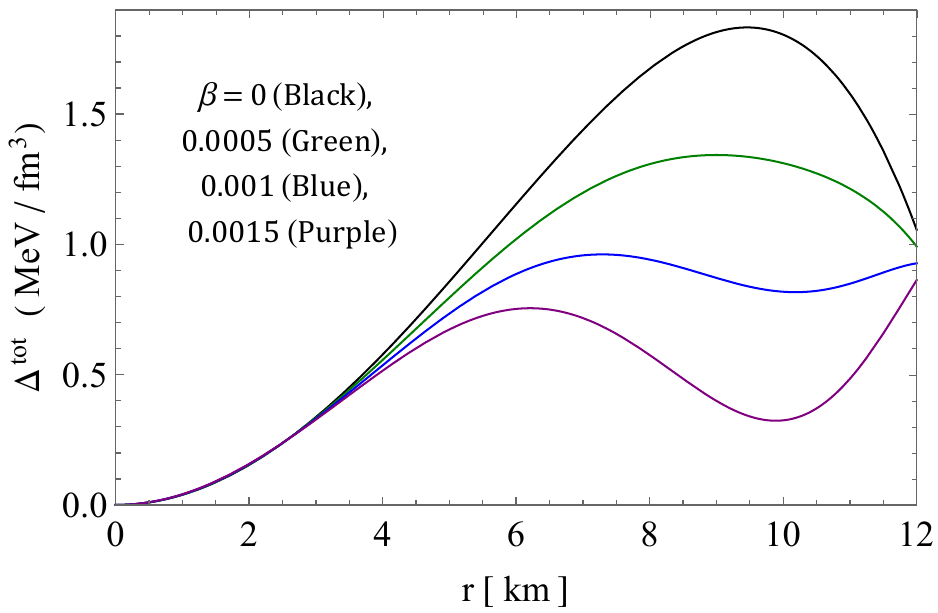}~~~~~
\includegraphics[height=6cm,width=7cm]{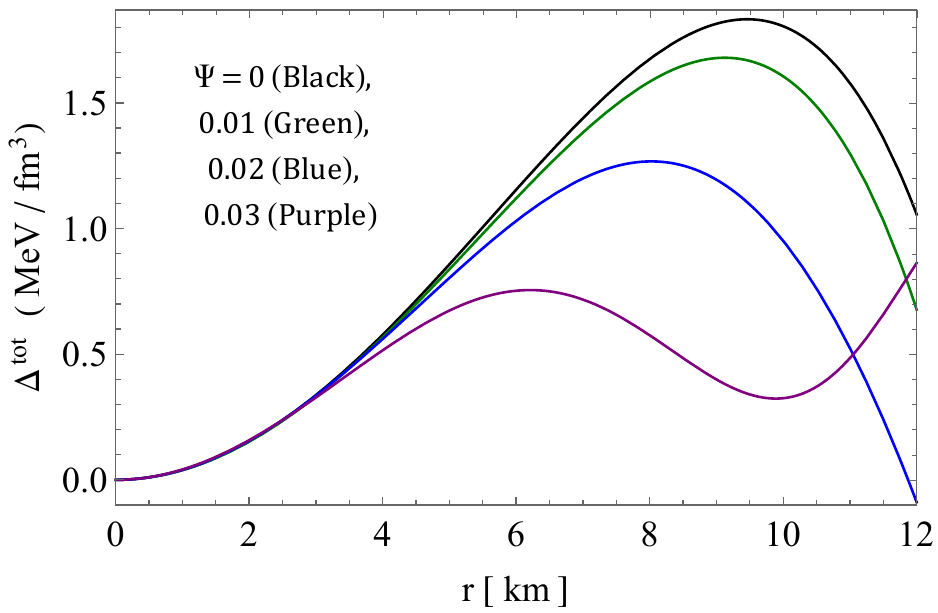}
\caption{Pressure anisotropy against radial coordinate $r$ for different values of $\beta$ and $\Psi$ with $r_s =12\,km,\,\rho_0 =0.00025/km^2, \, \rho_s=0.00020/km^2,\, \Psi=0.03/km^2$ (left panel) and  $r_s =12\,km,\,\rho_0 =0.00025/km^2, \, \rho_s=0.00020/km^2,\, \beta=0.0015$ (right panel) respectively.}
\label{f3}
\end{figure*}

Figure \ref{f3} depicts the anisotropic behavior defined by $\Delta^{\text{tot}} = P_t^{\text{tot}} - P_r^{\text{tot}}$, throughout the stellar interior for various values of the deformation parameter $\beta$ and the perturbation parameter $\Psi$. The profiles indicate that $\Delta^{\text{tot}}$ stays positive throughout the star, signifying that $P_t^{\text{tot}}$ continuously beats $P_r^{\text{tot}}$. This positive pressure difference act as an additional outward force that helps in counterbalancing of the gravitational force. Such kind of force is more essential for compact stars such as SSs, where the gravitational force is too strong. 
For lower values of $\beta$ and $\Psi$, anisotropy progressively rises with radius before stabilizing, whereas higher values produce more complicated patterns with small oscillations.

\subsection{Implications of perturbed functions and density constraints on $M-R$ relation profiles}

\begin{figure*}[!htp]
\centering
\includegraphics[height=6cm,width=7cm]{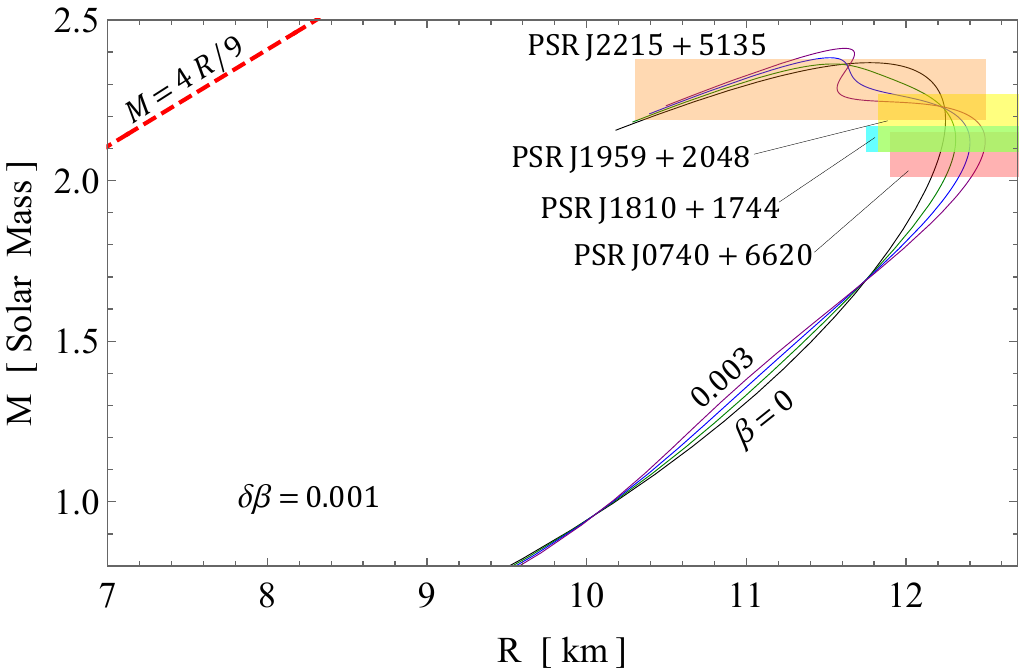}~~~~~
\includegraphics[height=6cm,width=7cm]{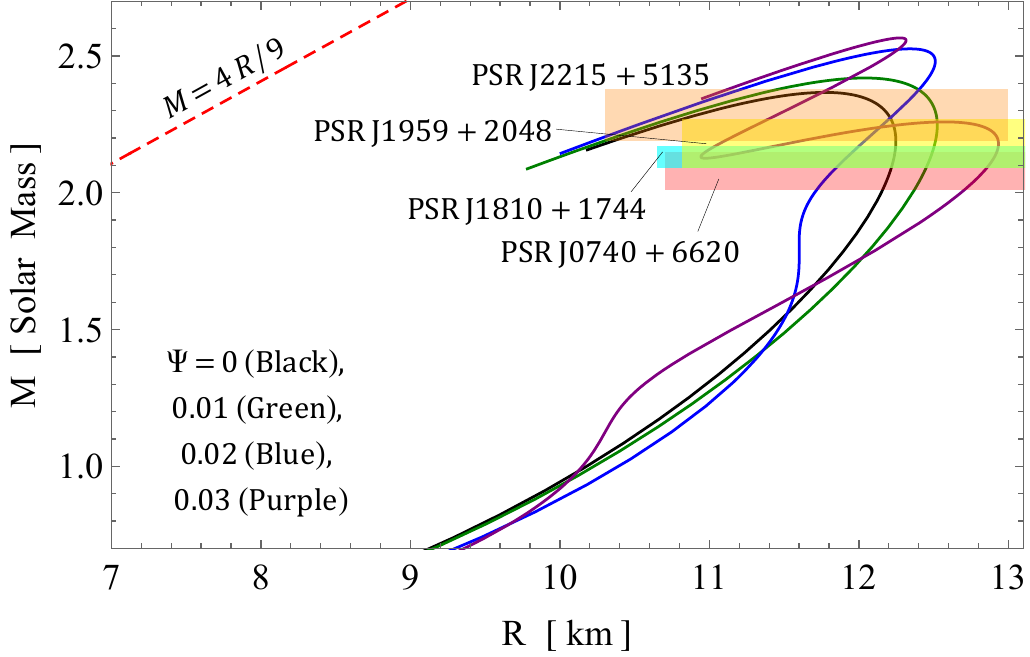}
\caption{Variations of mass with radius for different $\beta$ ($\Psi =0.03/km^2$) and $\Psi$ ($\beta =0.0075$) with $r_s =12\,km,\,\rho_0 =0.00025/km^2, \, \rho_s=0.00020/km^2$.}
\label{f4}
\end{figure*}

We examined the $M-R$ relationship for our stellar model, as seen in Fig.~\ref{f4}. In the left panel, the perturbation parameter $\Psi$ remains constant while the deformation parameter $\beta$ is altered from $0$ to $0.003$. The findings show a distinct pattern as raising $\beta$ immediately increases the maximum mass and associated radius, indicating improved structural support, before causing a slow decrease at higher values. For higher $\beta$, minor variations occur in the $M-R$ curves. For instance, at $\beta = 0.003$, the greatest mass attains $M = 2.28^{+0.10}{-0.09},M\odot$ with a radius of $R = 11.57^{+0.87}_{-0.10},\text{km}$. As $\beta$ approaches zero, the model smoothly converges to the usual MIT Bag model equation of state, resulting in a reduced maximum mass without any strange phenomena.

This indicates that the deformation parameter $\beta$ is essential in strengthening the star's resilience against gravitational collapse while maintaining physical consistency. Consequently, it allows the model to reach the mass-gap region, situated between the heaviest known neutron stars and the lightest stellar-mass black holes. Notably, for values of $\beta$ between $0$ and $0.003$, the model incorporates many precisely measured high-mass NeStrs. These include PSR J0740+6620 \cite{NANOGrav:2019jur,Fonseca:2021wxt}, with a mass of $2.08^{+0.07}_{-0.07}\,M_\odot$; PSR J1810+1744 \cite{PSRJ1810+1744}, at $2.13^{+0.04}_{-0.04}\,M_\odot$; PSR J1959+2048 \cite{starPSR}, with $2.18^{+0.09}_{-0.09}\,M_\odot$; and PSR J2215+5135 \cite{starPSR}, estimated at $2.28^{+0.10}_{-0.09}\,M_\odot$. These findings demonstrate the physical significance of the deformation parameter $\beta$, revealing its role in enhancing structural stability and offering a plausible explanation for the presence of exceptionally massive compact stars, potentially overcoming the theoretical gap between typical Neutron Stars and more exotic ultradense objects.

\begin{figure*}[!htp]
\centering
\includegraphics[height=6cm,width=7cm]{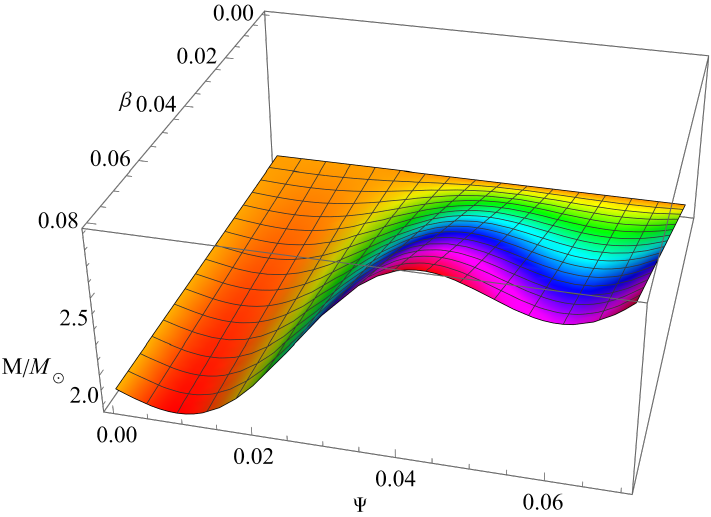}~~~~~
\includegraphics[height=6cm,width=7cm]{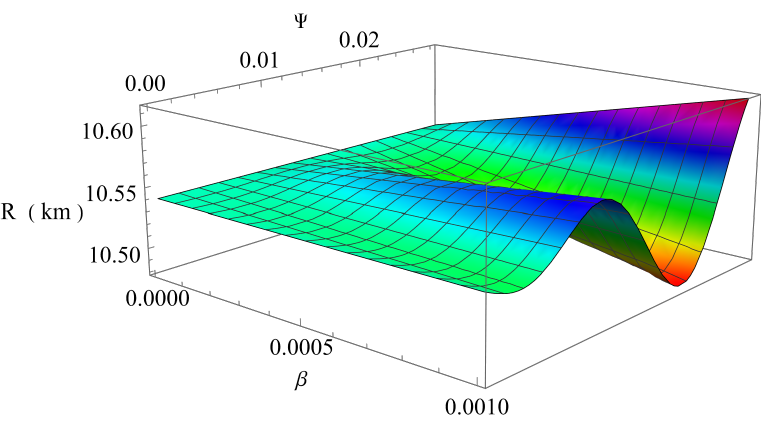}
\caption{Variations of mass and radius with respect to $\beta$ and $\Psi$ with $r_s =12\,km,\,\rho_0 =0.00025/km^2, \, \rho_s=0.00020/km^2$.}
\label{f5}
\end{figure*}

\begin{table*}[!htp]
\centering
\caption{The predicted radii of few high mass compact stars (from Fig. \ref{f4}).}\label{tab1}
\begin{minipage}{0.6\textheight}
 \scalebox{0.6}{\begin{tabular}{| *{10}{c|} }
\hline
 &    & \multicolumn{8}{c|}{Predicted $R$ (km)}  \\[0.15cm]
\cline{3-10}
{Objects} & {$\frac{M}{M_\odot}$} & \multicolumn{4}{c|}{ $\beta$} & \multicolumn{4}{c|}{$\Psi$} \\[0.15cm]
\cline{3-10}
&  & $0$ & $0.001$ & $0.002$ & $0.003$ & $0$ & $0.01$ & $0.02$ & $0.03$  \\[0.15cm] \hline
PSR J074+6620 \cite{NANOGrav:2019jur,Fonseca:2021wxt} & 2.08$\pm$0.07  & $12.21_{-0.05}^{+0.03}$  &  $12.30_{-0.05}^{+0.01}$  &   $12.39_{-0.06}^{+0.01}$  &  $12.48_{-0.07}^{+0.01}$  &  $12.21_{-0.05}^{+0.04}$  &   $12.45_{-0.08}^{+0.04}$  &  $\underbar{11.84}_{-0.12}^{+0.12}$  &   $\underbar{12.84}_{-0.13}^{+0.09}$     \\[0.15cm]
\hline
PSR J1810+1744 \cite{PSRJ1810+1744} & 2.13$\pm$0.04  & $12.24_{-0.02}^{+0.01}$  &  $12.31_{-0.07}^{+0.01}$  &   $12.40_{-0.07}^{+0.01}$  &  $12.50_{-0.09}^{+0.01}$  &  $12.24_{-0.02}^{+0.01}$  &   $12.48_{-0.11}^{+0.02}$  &  $11.92_{-0.07}^{+0.08}$  &   $12.91_{-0.05}^{+0.03}$    \\[0.15cm]
\hline
PSR J1959+2048 \cite{starPSR} & 2.18$\pm$0.09  & $12.25_{-0.03}^{+0.04}$  &  $12.29_{-0.17}^{+0.01}$  &   $12.36_{-0.44}^{+0.03}$  &  $12.45_{-}^{+0.04}$  &  $12.25_{-0.03}^{+0.01}$  &   $12.51_{-0.06}^{+0.01}$  &  $\underbar{12.02}_{-0.17}^{+0.18}$  &   $\underbar{12.94}_{-0.08}^{-}$   \\[0.15cm]
\hline
PSR J2215+5135 \cite{starPSR} & $2.28^{+0.10}_{-0.09}$ &  $12.20_{-}^{+0.05}$  &  $12.08_{-}^{+0.20}$  &   $\underbar{11.82}_{-0.24}^{+0.52}$  &  $\underbar{11.57}_{-0.10}^{+0.87}$  &  $12.21_{-}^{+0.04}$  &   $12.52_{-0.23}^{+0.01}$  &  $\underbar{12.23}_{-0.20}^{+0.18}$  &   $\underbar{11.43}_{-0.32}^{+0.37}$    \\[0.15cm]
\hline
\end{tabular}}
\end{minipage}
\\ Underlined ($\_$) values belong to perturbed ones
\end{table*}

The right panel of Fig. \ref{f4} exhibits a trend similar to that seen with changes in the deformation parameter $\beta$, though with small differences. In this context, we set $\beta$ constant and investigate the impact of the perturbation parameter $\Psi$ on the $M-R$ relation. The perturbation parameter $\Psi$ is modified within the interval $0 \leq \Psi \leq 0.03$ for this investigation. The data indicate a distinct trend as when $\Psi$ increases, both the maximum mass and radius initially increase, signifying strengthened structural support. Beyond a specific limit, this trend inversely shifts, resulting in a steady decrease in mass and radius, accompanied by increasingly significant fluctuations. The fluctuations are particularly evident at $\Psi \gtrsim 0.02$, suggesting that more intense disturbances substantially influence the dynamics of the fundamental strange matter. This interaction effectively mitigates the equation of state (EOS), constraining the star's maximum mass ($M_{\text{max}}$) and proving it progressively more challenging for such objects to collapse into low-mass black holes. The amplitude of $\Psi$ dictates this limiting effect as greater perturbations enhance internal structural fluctuations, enhancing the star's motion while simultaneously reducing the maximum sustainable mass. 

Within the considered range, $0 \leq \Psi \leq 0.03$, the model remains compatible with observations of high-mass pulsars: PSR J0740+6620~\cite{NANOGrav:2019jur,Fonseca:2021wxt} ($M = 2.08^{+0.07}_{-0.07},M_\odot$), PSR J1810+1744~\cite{PSRJ1810+1744} ($M = 2.13^{+0.04}_{-0.04},M_\odot$), PSR J1959+2048~\cite{starPSR} ($M = 2.18^{+0.09}_{-0.09},M_\odot$), and PSR J2215+5135~\cite{starPSR} ($M = 2.28^{+0.10}_{-0.09},M_\odot$), which are indicated in Fig.~\ref{f5} by horizontal bands for reference. These findings highlight the physical significance of $\Psi$: it not only influences internal stability and balance but also imposes a physically meaningful limit on $M_{\text{max}}$, making it a key factor in reconciling theoretical models with the existence of unusually massive compact stars.

Fig.~\ref{f5} further illustrates the star's response to perturbations. On the left, the mass---starting at $2,M_\odot$---shows minor oscillations under small $\beta$ and $\Psi$, but the fluctuations grow as both parameters increase. The radius, initially set at $12,\text{km}$, behaves similarly: small perturbations produce almost no change, whereas larger $\beta$ and $\Psi$ values induce significant swings. These amplified variations reflect a more reactive internal structure, sensitive to internal pressures and instabilities. Collectively, these results show that $\beta$ and $\Psi$ act as dynamic levers shaping the star's behavior, governing the complex interplay between deformation, perturbation, and equilibrium in dense, exotic stellar objects.

\begin{figure*}
\centering
\includegraphics[height=6cm,width=7cm]{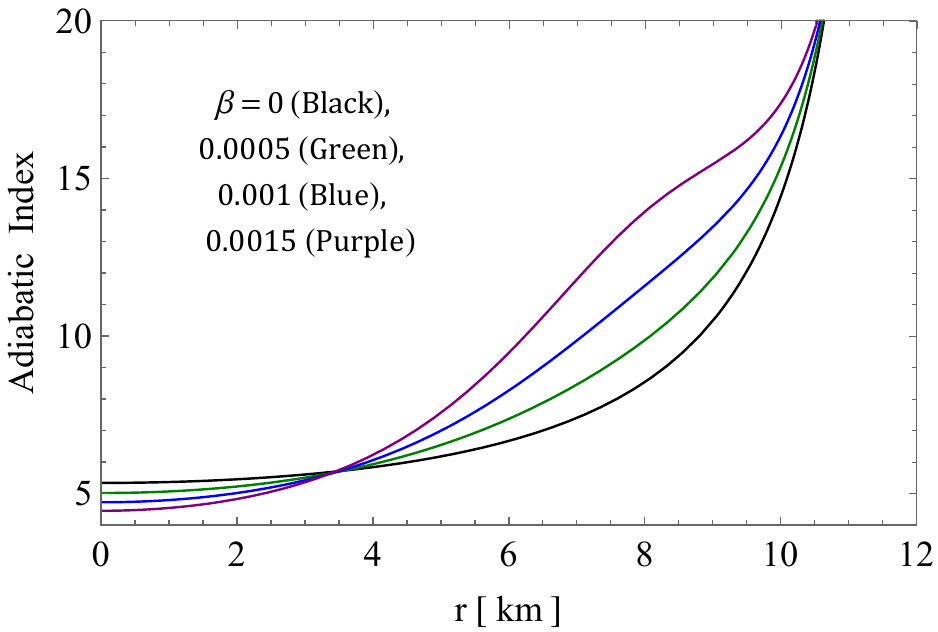}~~~~~
\includegraphics[height=6cm,width=7cm]{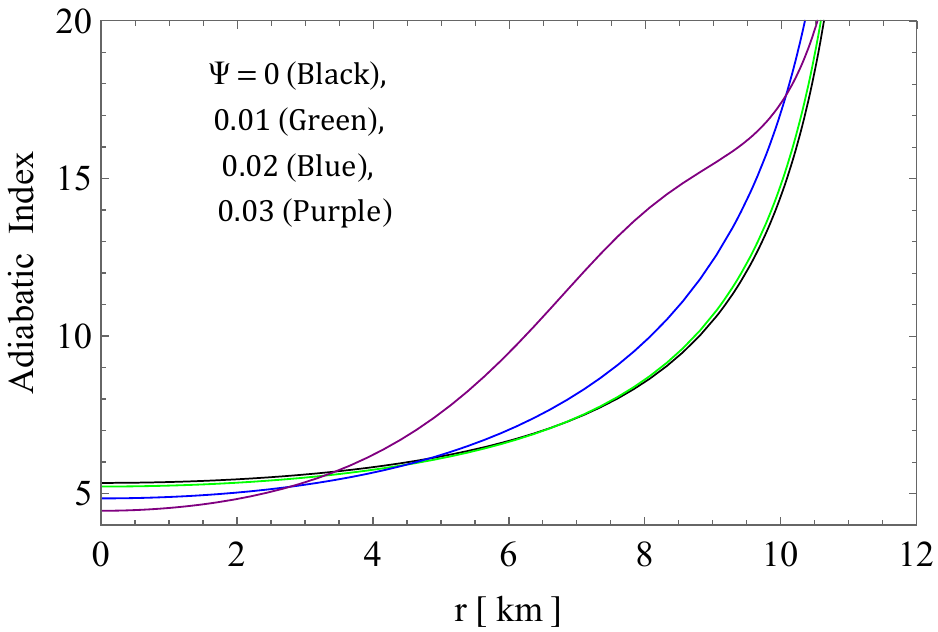}
\caption{Adiabatic indices against radial coordinate $r$ for different values of $\beta$ and $\Psi$ with $r_s =12\,km,\,\rho_0 =0.00025/km^2, \, \rho_s=0.00020/km^2,\, \Psi=0.03/km^2$ (left panel) and  $r_s =12\,km,\,\rho_0 =0.00025/km^2, \, \rho_s=0.00020/km^2,\, \beta=0.0015$ (right panel) respectively.}
\label{f6}
\end{figure*}

\section{Implications of Perturbed functions on Stability of the Model}\label{sec6}

\subsection{Stability Analysis via Adiabatic Index} 

It is now necessary to assess the stability of the perturbed stellar structure. This necessitates an examination of the adiabatic index ($\Gamma$), which is indicated by the subsequent expression:
\begin{equation}
    \Gamma = \frac{\rho^{\text{tot}}+P^{\text{tot}}_r}{P^{\text{tot}}_r}~\frac{dP^{\text{tot}}_r}{d\rho^{\text{tot}}}.
\end{equation}
In the Newtonian framework, the stability condition for an isotropic fluid is expressed as $\Gamma > \frac{4}{3}$.  Heintzmann and Bondi~\cite{heintzmann1975neutron,bondi1992anisotropic} emphasize this criterion in their research on anisotropic fluids and NeStrs.  Conversely, the stability condition for an anisotropic stellar model deviates from the conventional Chandrasekhar result that is applicable to isotropic fluids \cite{chan1992dynamical,chan1993dynamical}. This deviation can be understand by the following equation, 
\begin{equation}
\Gamma > \frac{4}{3}\left(1 + \frac{\Delta^\text{tot}}{r|(P^{\text{tot}}_{r})|^{\prime}}+\frac{1}{4}\frac{8\pi \rho^{\text{tot}} P^{\text{tot}}_{r}r}{|(P^{\text{tot}}_{r})|^{\prime}} \right).\label{eq62}
\end{equation}

In this scenario, primes signify differentiation within a radial coordinate, represented as $r$. The second component in Eq. (\ref{eq62}) pertains to the alterations in the stability condition due to anisotropy ($P^{\text{tot}}_r \neq P^{\text{tot}}_t$), while the last term includes relativistic corrections. 

The adiabatic index, represented as $\Gamma$, is constrained by certain conditions concerning the dynamical instability of an isotropic fluid sphere. The crucial limit is known as the critical adiabatic index, $\Gamma_{cr}$, which refers to the inequality, 
\begin{equation}
\langle \Gamma \rangle > \Gamma_{cr},
\end{equation}
here, the average adiabatic index is shown by $\langle \Gamma \rangle$ \cite{Moustakidis:2016ndw}. 

As shown in references \cite{Moustakidis:2016ndw, Koliogiannis:2018hoh}, the critical value may be written as  
\begin{equation}
\Gamma_{cr} = \frac{4}{3} + \frac{19}{42} C,
\end{equation}
where $C = \frac{2M}{R}$ denotes the compactness ratio in the context of GR. Within the framework of pure Newtonian gravity, the critical value is invariant at $\Gamma_{cr} = \frac{4}{3}$. Nevertheless, after accounting for the effects of GR, this number exceeds $\frac{4}{3}$. In Fig.~\ref{f6}, we illustrate the behavior of the adiabatic index $\Gamma$, which displays a pronounced sensitivity to variations in both the deformation parameter $\beta$ and the perturbation parameter $\Psi$. As $\beta$ is increased from 0 to 0.0015 and $\Psi$ from 0 to $0.03\,\text{km}^{-2}$, the radial profile of $\Gamma$ reveals increasingly distinct deviations across the stellar interior. Although both $\beta$ and $\Psi$ affect the adiabatic index, the variations remain modest, indicating that the stellar matter responds stably to these perturbations. The central value of $\Gamma$ approaches the critical $4/3$ threshold more closely as $\beta$ and $\Psi$ increase, signaling enhanced resistance to radial instability and greater stability under extreme densities---an important characteristic for compact star models. At the core, higher $\beta$ and $\Psi$ slightly reduce the effective quark pressure, softening the EOS in that region. Toward the surface, the opposite occurs: the matter becomes stiffer, producing a contrasting behavior between the inner and outer layers. This highlights the complex internal dynamics induced by these parameters. Overall, Fig.~\ref{f5} shows that the anisotropic stellar configuration maintains hydrostatic stability across the considered range of $\beta$ and $\Psi$, underscoring the physical consistency and astrophysical relevance of the model for describing realistic relativistic compact stars.

\subsection{Stability analysis via speed of sound criterion}
Understanding sound propagation within a compact star is crucial for evaluating the physical validity of the stellar model. This propagation is determined by the relationship $\text{v}^2_s = \frac{dP}{d\rho}$, which incorporates the sensitivity of pressure to variations in energy density. A crucial need of any realistic physical model is to maintain the principle of causality, specifically that no signal, including sound, may propagate at a velocity exceeding the speed of light. In natural units, where the speed of light is normalized to one ($c = 1$), this condition simply corresponds to the constraint $\text{v}^2_s < 1$ within the stellar interior. To investigate the interior dynamics of a compact anisotropic star, it is essential to figure out the propagation of perturbations, such as sound waves, across its matter distribution. Anisotropy induces directional variation in pressure, resulting in the star's different response to disturbances along radial and tangential directions. This necessitates the separate analysis of sound propagation along each direction, yielding two distinct sound speed profiles.

The radial and tangential sound speeds are defined as
\begin{eqnarray}
\text{v}^2_r = \frac{dP^{\text{tot}}_r}{d\rho}~~,~~~~\text{v}^2_t = \frac{dP^{\text{tot}}_t}{d\rho},
\end{eqnarray}
which assess the response of radial and tangential pressures to variations in energy density. These velocities offer critical insights into the star's local mechanical dynamics and act as indications of dynamical stability. By verifying that both $\text{v}^2_r$ and $\text{v}^2_t$ stay lower to the speed of light throughout the star which guarantee that the model adheres to causality and constitutes a physically consistent framework for characterizing dense, relativistic compact objects such as SSs.

\begin{figure*}[!htp]
\centering
\includegraphics[height=6cm,width=7cm]{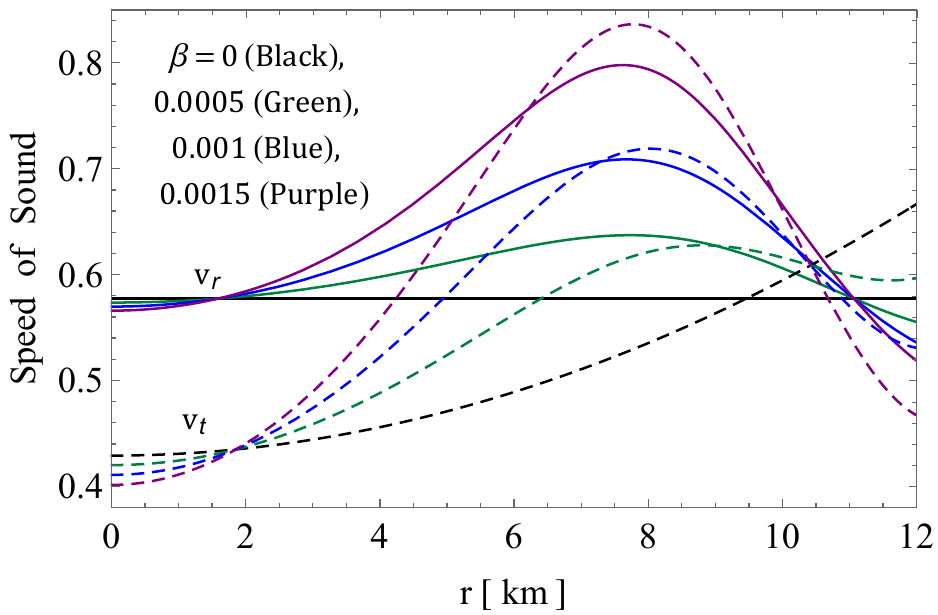}~~~~~
\includegraphics[height=6cm,width=7cm]{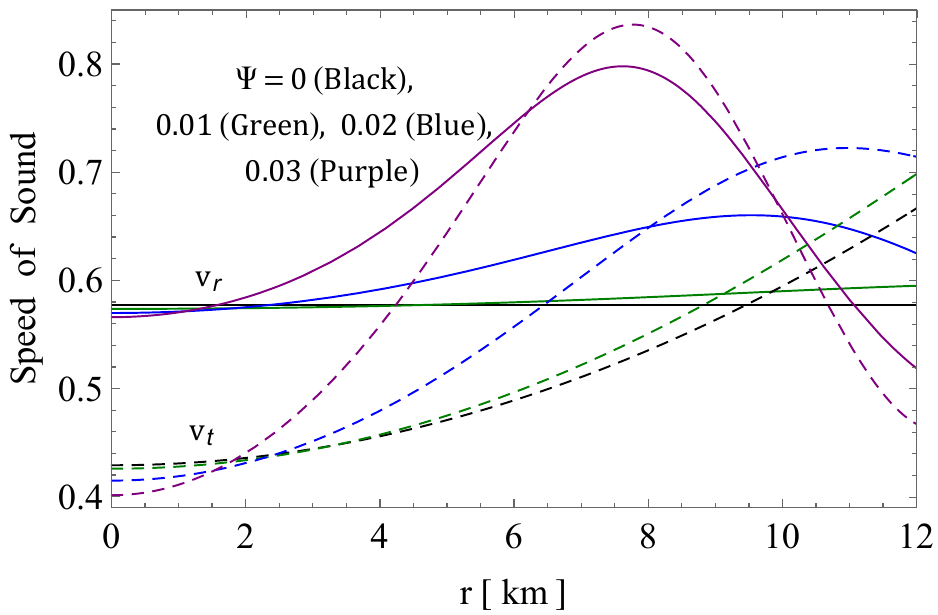}
\caption{Speed of sounds against radial coordinate $r$ for different values of $\beta$ and $\Psi$ with $r_s =12\,km,\,\rho_0 =0.00025/km^2, \, \rho_s=0.00020/km^2,\, \Psi=0.03/km^2$ (left panel) and  $r_s =12\,km,\,\rho_0 =0.00025/km^2, \, \rho_s=0.00020/km^2,\, \beta=0.0015$ (right panel) respectively.}
\label{f7}
\end{figure*}

Figure~\ref{f7} analyzes the behavior of sound speeds within the SS as the deformation parameter $\beta$ and the perturbation magnitude $\Psi$ are varied. The radial sound speed $\text{v}^2_r$ and the tangential sound speed $\text{v}^2_t$ progressively vary from the core  to boundary which shows that it never surpasses the speed of light, and hence maintaining the model's physical consistency. The tangential sound speed $\text{v}^2_t$ exhibits significant sensitivity; as $\beta$ and $\Psi$ grow, it rises from approximately 0.40 to roughly 0.84, indicating a major influence of these parameters on the star's internal pressure response to perturbations. Nonetheless, substantially modifying either parameter results in $\text{v}^2_r$ beyond unity, hence breaching causality and rendering the model unphysical. Consequently, the sound speeds not only characterize the star's causality but also serve as a natural barrier, indicating the thresholds beyond which the model fails to accurately depict true stellar behavior.  It keeps us grounded by showing us not only how the model works but also where its real limits lies. For the current model, the principle of causality holds for $\beta \in [0,\,0.002]$ and $\Psi \in [0,\,0.035]$.

\subsection{Stability analysis via Harrison-Zel'dovich-Novikov criteria} 
\begin{figure*}[!htp]
\centering
\includegraphics[height=6cm,width=7cm]{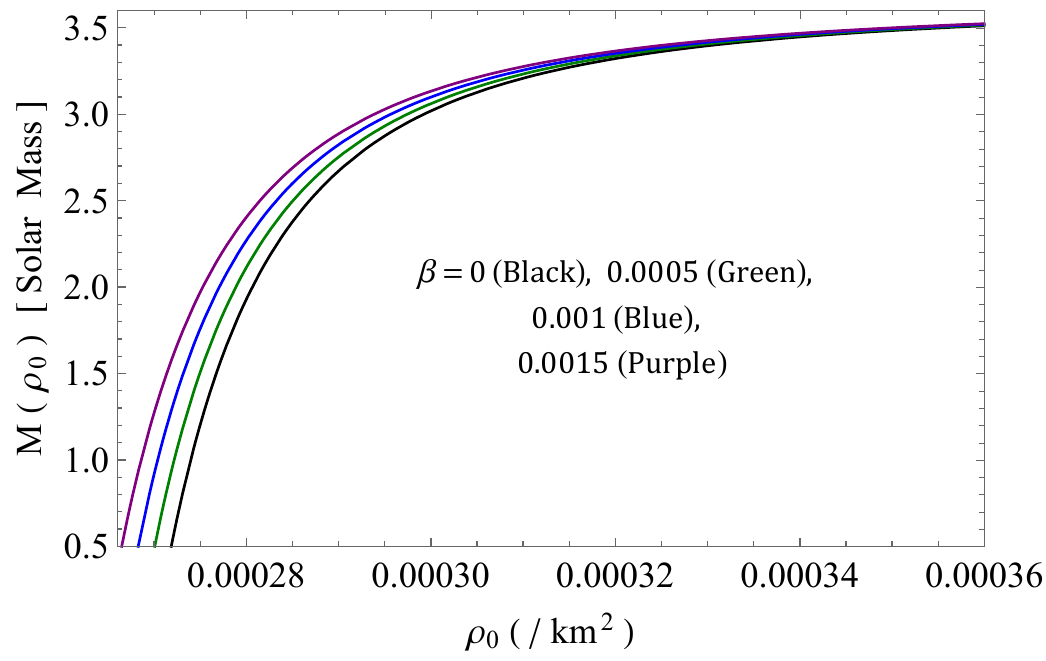}~~~~~
\includegraphics[height=6cm,width=7cm]{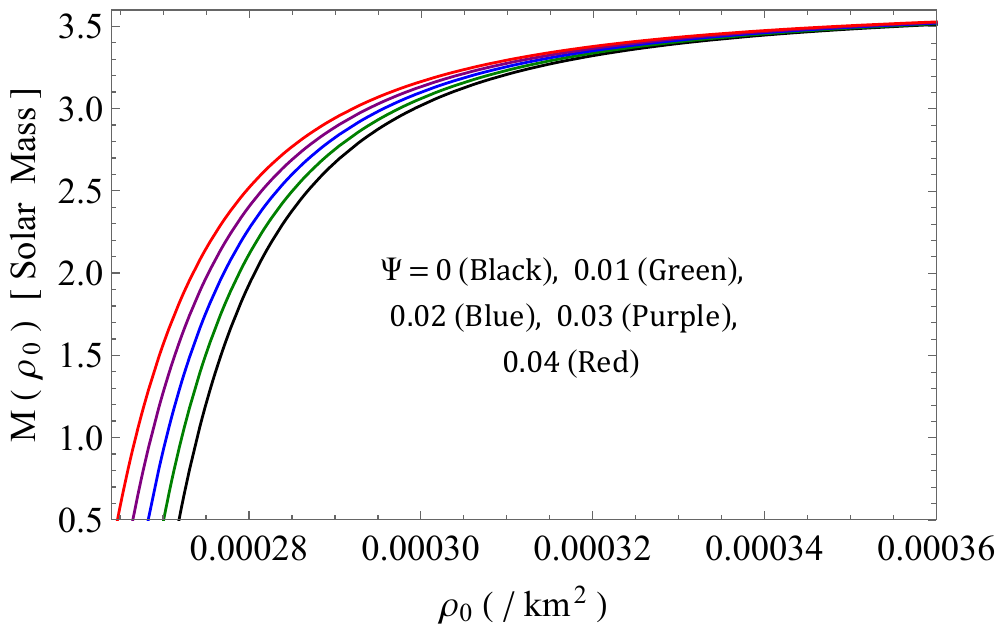}
\caption{Mass ($M/M_\odot$) versus central density $(\rho_0)$ for different values of $\beta$ and $\Psi$ with $\rho_0 =0.00025/km^2, \, \rho_s=0.00020/km^2,\, \Psi=0.03/km^2$ (left panel) and  $r_s =12\,km,\,\rho_0 =0.00025/km^2, \, \rho_s=0.00020/km^2,\, \beta=0.0015$ (right panel) respectively.}
\label{f8}
\end{figure*}

For the perturbed solution, we will now investigate the connection between stellar mass $M$ and central energy density $\rho^{\text{tot}}_0$. In order to do this, the following static stability criteria is used \cite{ZHN1, ZHN2}, 
 \begin{itemize}
\item~~~~ $dM/d\,\rho^{\text{tot}}_0 < 0 \hspace{0.5cm} \rightarrow \mbox{unstable configuration}$,\\
 \item ~~~~  $dM/d\, \rho^{\text{tot}}_0 > 0 \hspace{0.5cm} \rightarrow \mbox{stable configuration}$,
\end{itemize}
The following formulae describe the mass equation as an analytic function of $\rho_0$, which allows us to prove the aforementioned requirement for perturbed solution:
\begin{small}
\begin{eqnarray}
\hspace{-0.5cm} M=\frac{8 \pi  r_s^2 \left(5 \rho^{\text{tot}}_0 r_s^2+3 \rho^{\text{tot}}_s r_s^2 -3 \rho^{\text{tot}}_0 r_s^2\right)}{15 r_s^2}-\beta  \sin \left(r_s^2 \Psi \right), 
\end{eqnarray}
\end{small}
A positive derivative $\frac{dM}{d\rho^{\text{tot}}_0}$ is expected, as illustrated in Fig.~\ref{f8}, where $M(\rho^{\text{tot}}_0)$ increases steadily. This confirms that the perturbed model remains stable. Introducing the perturbation function results in a larger mass increase across the same range of central densities $\rho^{\text{tot}}_0$, indicating that the perturbation can enhance the stability of the compact star when its amplitude and frequency are appropriately tuned.

\section{Concluding remarks}\label{sec7} 

In this study, we derived an exact analytical solution to the Einstein field equations \eqref{eq9}-\eqref{eq11}, describing the internal structure of a SS within standard GR. Together with the Tolman-Oppenheimer-Volkoff equation \eqref{eq7}, these equations govern how gravity, pressure, and anisotropic forces balance inside a dense compact object. Analytical solutions are difficult to obtain because of their nonlinear characteristics and realistic matter composition. Therefore, we employed a physically motivated perturbative method to account for minor, stable perturbations that strange stars may encounter in astrophysical contexts, including low-level accretion, subtle magnetic changes, or transient gravitational waves.

We implemented a sinusoidal perturbation in the metric functions represented by $g(r) = \sin(\Psi r^2)$ to reflect these effects. This generates a uniform, symmetrical deformation of the radial coordinate. This selection guarantees mathematical accessibility, physical consistency at the star core, and simulates natural quasi-normal oscillations. The impact of minimal spacetime distortions on internal structure may be analyzed by incorporating this perturbation into the field equations. The resultant anisotropic stresses, $\Delta^{\text{tot}} = P_t^{\text{tot}} - P_r^{\text{tot}}$, indicate the internal tension of the stellar fluid. It describes its capacity for resisting collapse, especially in relation to the outer layers. Explicit expressions for the constants $\mathcal{B}_g$ and $\mathcal{F}$ were derived in relation to $\beta$, $\Psi$, and additional theoretical parameters by aligning the interior solution with the Schwarzschild exterior at $r=R$ through the Israel-Darmois junction conditions and imposing zero radial pressure at the surface. This approach facilitates the examination of how minor geometric disturbances alter the star's interior structure while preserving mass and radius in accordance with observations. 

Thermodynamic analysis shows that energy density $\rho$ and pressures $P_r$ and $P_t$ remain positive, finite, and smoothly decreasing from the core to the surface for all tested values of $\beta$ and $\Psi$. This confirms both physical consistency and dynamic stability. The model aligns with high-mass pulsar observations, including PSR J0740+6620 ($2.08^{+0.07}_{-0.07}~M_\odot$), PSR J1810+1744 ($2.13^{+0.04}_{-0.04}~M_\odot$), PSR J1959+2048 ($2.18^{+0.09}_{-0.09}~M_\odot$), and PSR J2215+5135 ($2.28^{+0.10}_{-0.09}~M_\odot$). The anisotropy profile remains positive throughout the star, gradually increasing for small $\beta$ and $\Psi$. However, it exhibits only mild oscillations at higher values, providing additional outward pressure. This strengthens structural stability, especially critical for SSs with extreme gravity.

Investigating the influence of $\beta$ and $\Psi$ on the $M-R$ relation demonstrates that an initial increase in these parameters improves the maximum mass and radius. For instance, $\beta = 0.003$ yields $M_{\text{max}} = 2.28^{+0.10}_{-0.09}~M_\odot$ and $R = 11.57^{+0.87}_{-0.10}~\text{km}$. Fluctuations occur above specific threshold values, especially for $\Psi \gtrsim 0.02$. This signifies a dynamic action within the star's internal structure. Dynamic studies, begin with $M = 2,M_\odot$ and $R = 12,\text{km}$, indicate that minimal $\beta$ and $\Psi$ yield negligible oscillations, whereas higher values induce the star to exhibit a "breathing" phenomenon. This simply indicates that mass and radius are oscillating more significantly. These findings emphasize that $\beta$ and $\Psi$ significantly influence star structure, evolution, and stability.

The stability analysis, that includes the adiabatic index $\Gamma$, sound velocity, and mass action, validates the model's robustness. $\Gamma$ consistently maintains below the critical value of $4/3$ everywhere, approaching it near to the core as $\beta$ and $\Psi$ increase, signifying enhanced resistance to collapse. The radial and tangential sound speeds remain subluminal, thereby maintaining causality, however $\text{v}_t^2$ increases from around 0.40 to 0.84 with greater perturbations, indicating a pronounced sensitivity of internal pressure to these parameters. Excessive $\beta$ or $\Psi$ would violate causality, establishing a distinct physical boundary. The mass increases consistently with central energy density, and the derivative $dM/d\rho_0$ remains positive which introduce a perturbation increases total mass within the same density range, suggesting that controlled disturbances may strengthen stability.

In general, the model exhibits physical and dynamic consistency throughout the tested ranges of $\beta$ and $\Psi$. The internal equilibrium of SSs is significantly altered by even minor deformations and perturbations, indicating that delicate, ongoing environmental influences can leave lasting effects on their structure. This method offers a robust framework for the investigation of realistic, relativistic compact stars in perturbed conditions.

\section*{Acknowledgments}
This research was funded by the Science Committee of the Ministry of Science and Higher Education of the Republic of Kazakhstan (Grant No. AP23487178). This research was also funded by Institute of Eminence, University of Delhi (Ref. No./IoE/2025-26/12/FRP) through Faculty Research Program (FRP). SKM also thankful to UoN administration for continuous support and encouragement for the research works.  

\section*{Data Availability Statement}
This manuscript has no associated data or the data will not be deposited, as such there is no observational data related to this article where the necessary calculations and graphic discussion are already available in the manuscript.

\section*{Conflict of Interest}
The authors declare that they have no conflict of interest or personal relationships that could have appeared to influence the work reported in this paper.


\begin{thebibliography}{99}
\bibitem{Stairs:2003eg}
I.~H.~Stairs,
Living Rev. Rel. \textbf{6}, 5 (2003)

\bibitem{Reardon:2015kba}
D.~J.~Reardon, G.~Hobbs, W.~Coles, Y.~Levin, M.~J.~Keith, M.~Bailes, N.~D.~R.~Bhat, S.~Burke-Spolaor, S.~Dai and M.~Kerr, \textit{et al.}
Mon. Not. Roy. Astron. Soc. \textbf{455}, no.2, 1751-1769 (2016)

\bibitem{Bogdanov:2019ixe}
S.~Bogdanov, S.~Guillot, P.~S.~Ray, M.~T.~Wolff, D.~Chakrabarty, W.~C.~G.~Ho, M.~Kerr, F.~K.~Lamb, A.~Lommen and R.~M.~Ludlam, \textit{et al.}
Astrophys. J. Lett. \textbf{887}, no.1, L25 (2019)

\bibitem{Bogdanov:2019qjb}
S.~Bogdanov, F.~K.~Lamb, S.~Mahmoodifar, M.~C.~Miller, S.~M.~Morsink, T.~E.~Riley, T.~E.~Strohmayer, A.~L.~Watts, A.~J.~Dittmann and D.~Chakrabarty, \textit{et al.}
Astrophys. J. Lett. \textbf{887}, no.1, L26 (2019)

\bibitem{LIGOScientific:2016aoc}
B.~P.~Abbott \textit{et al.} [LIGO Scientific and Virgo],
Phys. Rev. Lett. \textbf{116}, no.6, 061102 (2016)

\bibitem{Bhattacharyya:2016kte}
S.~Bhattacharyya, I.~Bombaci, D.~lnoteta and A.~V.~Thampan,
Mon. Not. Roy. Astron. Soc. \textbf{457}, no.3, 3101-3114 (2016)

\bibitem{Annala:2019puf}
E.~Annala, T.~Gorda, A.~Kurkela, J.~N{\"a}ttil{\"a} and A.~Vuorinen,
Nature Phys. \textbf{16}, no.9, 907-910 (2020)


\bibitem{Witten:1984rs}
E.~Witten,
Phys. Rev. D \textbf{30}, 272-285 (1984)

\bibitem{Farhi:1984qu}
E.~Farhi and R.~L.~Jaffe,
Phys. Rev. D \textbf{30}, 2379 (1984)

\bibitem{Ozel:2016oaf}
F.~{\"O}zel and P.~Freire,
Ann. Rev. Astron. Astrophys. \textbf{54}, 401-440 (2016)

\bibitem{NANOGrav:2019jur}
H.~T.~Cromartie \textit{et al.} [NANOGrav],
Nature Astron. \textbf{4}, no.1, 72-76 (2019)

\bibitem{Fonseca:2021wxt}
E.~Fonseca, H.~T.~Cromartie, T.~T.~Pennucci, P.~S.~Ray, A.~Y.~Kirichenko, S.~M.~Ransom, P.~B.~Demorest, I.~H.~Stairs, Z.~Arzoumanian and L.~Guillemot, \textit{et al.}
Astrophys. J. Lett. \textbf{915}, no.1, L12 (2021)

\bibitem{Miller:2021qha}
M.~C.~Miller, F.~K.~Lamb, A.~J.~Dittmann, S.~Bogdanov, Z.~Arzoumanian, K.~C.~Gendreau, S.~Guillot, W.~C.~G.~Ho, J.~M.~Lattimer and M.~Loewenstein, \textit{et al.}
Astrophys. J. Lett. \textbf{918}, no.2, L28 (2021)

\bibitem{Riley:2021pdl}
T.~E.~Riley, A.~L.~Watts, P.~S.~Ray, S.~Bogdanov, S.~Guillot, S.~M.~Morsink, A.~V.~Bilous, Z.~Arzoumanian, D.~Choudhury and J.~S.~Deneva, \textit{et al.}
Astrophys. J. Lett. \textbf{918}, no.2, L27 (2021)

\bibitem{Antoniadis:2013pzd}
J.~Antoniadis, P.~C.~C.~Freire, N.~Wex, T.~M.~Tauris, R.~S.~Lynch, M.~H.~van Kerkwijk, M.~Kramer, C.~Bassa, V.~S.~Dhillon and T.~Driebe, \textit{et al.}
Science \textbf{340}, 6131 (2013)


\bibitem{Demorest:2010bx}
P.~Demorest, T.~Pennucci, S.~Ransom, M.~Roberts and J.~Hessels,
Nature \textbf{467}, 1081-1083 (2010)


\bibitem{Fonseca:2016tux}
E.~Fonseca, T.~T.~Pennucci, J.~A.~Ellis, I.~H.~Stairs, D.~J.~Nice, S.~M.~Ransom, P.~B.~Demorest, Z.~Arzoumanian, K.~Crowter and T.~Dolch, \textit{et al.}
Astrophys. J. \textbf{832}, no.2, 167 (2016)

\bibitem{NANOGRAV:2018hou}
Z.~Arzoumanian \textit{et al.} [NANOGRAV],
Astrophys. J. \textbf{859}, no.1, 47 (2018)

\bibitem{Miller:2019cac}
M.~C.~Miller, F.~K.~Lamb, A.~J.~Dittmann, S.~Bogdanov, Z.~Arzoumanian, K.~C.~Gendreau, S.~Guillot, A.~K.~Harding, W.~C.~G.~Ho and J.~M.~Lattimer, \textit{et al.}
Astrophys. J. Lett. \textbf{887}, no.1, L24 (2019)

\bibitem{Raaijmakers:2019qny}
G.~Raaijmakers, T.~E.~Riley, A.~L.~Watts, S.~K.~Greif, S.~M.~Morsink, K.~Hebeler, A.~Schwenk, T.~Hinderer, S.~Nissanke and S.~Guillot, \textit{et al.}
Astrophys. J. Lett. \textbf{887}, no.1, L22 (2019)


\bibitem{Gonzalez-Caniulef:2019wzi}
D.~Gonzalez-Caniulef, S.~Guillot and A.~Reisenegger,
Mon. Not. Roy. Astron. Soc. \textbf{490}, no.4, 5848-5859 (2019)

\bibitem{LIGOScientific:2018cki}
B.~P.~Abbott \textit{et al.} [LIGO Scientific and Virgo],
Phys. Rev. Lett. \textbf{121}, no.16, 161101 (2018)

\bibitem{LIGOScientific:2020zkf}
R.~Abbott \textit{et al.} [LIGO Scientific and Virgo],
Astrophys. J. Lett. \textbf{896}, no.2, L44 (2020)


\bibitem{Doroshenko:2022nwp}
V.~Doroshenko, V.~Suleimanov, G.~P{\"u}hlhofer and A.~Santangelo,
Nature Astron. \textbf{6}, no.12, 1444-1451 (2022)


\bibitem{Romani:2022jhd}
R.~W.~Romani, D.~Kandel, A.~V.~Filippenko, T.~G.~Brink and W.~Zheng,
Astrophys. J. Lett. \textbf{934}, no.2, L17 (2022)

\bibitem{ElHanafy:2023vig}
W.~El Hanafy and A.~Awad,
Astrophys. J. \textbf{951}, no.2, 144 (2023)

\bibitem{Yang:2020xyi}
Y.~Yang, V.~Gayathri, I.~Bartos, Z.~Haiman, M.~Safarzadeh and H.~Tagawa,
Astrophys. J. Lett. \textbf{901}, no.2, L34 (2020)

\bibitem{Alho:2022bki}
A.~Alho, J.~Nat{\'a}rio, P.~Pani and G.~Raposo,
Phys. Rev. D \textbf{106}, no.4, L041502 (2022)

\bibitem{Alho:2021sli}
A.~Alho, J.~Nat{\'a}rio, P.~Pani and G.~Raposo,
Phys. Rev. D \textbf{105}, no.4, 044025 (2022)
[erratum: Phys. Rev. D \textbf{105}, no.12, 129903 (2022)]

\bibitem{Roupas:2020mvs}
Z.~Roupas and G.~G.~L.~Nashed,
Eur. Phys. J. C \textbf{80}, no.10, 905 (2020)

\bibitem{Raposo:2018rjn}
G.~Raposo, P.~Pani, M.~Bezares, C.~Palenzuela and V.~Cardoso,
Phys. Rev. D \textbf{99}, no.10, 104072 (2019)

\bibitem{Cardoso:2019rvt}
V.~Cardoso and P.~Pani,
Living Rev. Rel. \textbf{22}, no.1, 4 (2019)

\bibitem{Errehymy:2024tqr}
A.~Errehymy, I.~Karar, K.~Myrzakulov, A.~Banerjee, A.~H.~Abdel-Aty and K.~S.~Nisar,
JHEAp \textbf{44}, 410-418 (2024)


\bibitem{Maurya:2024bfw}
S.~K.~Maurya, A.~Errehymy, B.~Dayanandan, O.~Donmez, K.~Myrzakulov, K.~S.~Nisar and M.~Mahmoud,
JHEAp \textbf{45}, 46-61 (2025)


\bibitem{Maurya:2024zao}
S.~K.~Maurya, M.~K.~Jasim, A.~Errehymy, K.~Boshkayev, G.~Mustafa and B.~Dayanandan,
Phys. Dark Univ. \textbf{46}, 101665 (2024)

\bibitem{Maurya:2024ylr}
S.~K.~Maurya, A.~Errehymy, K.~Newton Singh, A.~Aziz, S.~Hansraj and S.~Ray,
Astrophys. J. \textbf{972}, no.2, 175 (2024)

\bibitem{Hansraj:2024cgw}
S.~Hansraj and A.~Errehymy,
Phys. Dark Univ. \textbf{46}, 101632 (2024)


\bibitem{IIbragimov1}
B. Rahmatov, S. Murodov, J. Rayimbaev, Y. Turaev, I. Egamberdiev, Badalov, S. Ahmedov, S. Usanov
Ann. Phys.,\textbf{488} 171366 (2026)

\bibitem{IIbragimov2} 
B. Rahmatov, I.Egamberdiev. O. Umarov, M. Vapayev, S. Karshiboev, Y. Turaev, S. Murodov 
Nucl. Phys. B \textbf{1022} 117212 (2026)

\bibitem{IIbragimov3} 
S. Khan, J. Rayimbaev, S. Iskandarov, A. Seytov, I Ibragimov, S. Muminov, 
Phys. Dark Universe \textbf{51}  102220 (2026)

\bibitem{Malik:2024tto}
A.~Malik, A.~Almas, A.~Shafaq and F.~Mofarreh,
Int. J. Geom. Meth. Mod. Phys. \textbf{22}, no.05, 2450331 (2025)

\bibitem{Varela:2010mf}
V.~Varela, F.~Rahaman, S.~Ray, K.~Chakraborty and M.~Kalam,
Phys. Rev. D \textbf{82}, 044052 (2010)

\bibitem{Malik:2024boe}
A.~Malik, M.~Batool, E.~Meer, M.~F.~Shamir and A.~H.~Alkhaldi,
Int. J. Geom. Meth. Mod. Phys. \textbf{21}, no.10, 2440025 (2024)

\bibitem{Ashraf:2024cww}
A.~Ashraf, F.~Javed, W.~X.~Ma and G.~Mustafa,
Int. J. Geom. Meth. Mod. Phys. \textbf{21}, no.09, 2450161 (2024)

\bibitem{Maurya:2021aio}
S.~K.~Maurya, F.~Tello-Ortiz and S.~Ray,
Phys. Dark Univ. \textbf{31}, 100753 (2021)

\bibitem{Ditta:2023huk}
A.~Ditta, X.~Tiecheng, M.~Asia and I.~Mahmood,
Int. J. Geom. Meth. Mod. Phys. \textbf{21}, no.04, 2450076 (2024)

\bibitem{Maurya:2023uiy}
S.~K.~Maurya, G.~Mustafa, S.~Ray, B.~Dayanandan, A.~Aziz and A.~Errehymy,
Phys. Dark Univ. \textbf{42}, 101284 (2023)

\bibitem{Rani:2023vha}
S.~Rani, M.~Adeel, M.~Z.~Gul and A.~Jawad,
Int. J. Geom. Meth. Mod. Phys. \textbf{21}, no.01, 2450033 (2024)



\bibitem{Kumar:2025flr}
J.~Kumar, G.~Miglani, S.~Chaudhary and R.~S.~Chandelkar,
Phys. Scripta \textbf{100}, no.3, 035015 (2025)


\bibitem{Zoya:2026npb}
Z.~Asghara, M.~Farasat~Shamir, F.~Mofarreh, J.~Rayimbaev, O.~Sirajiddin, F.~Shayimov,
Nuc. Phys B. \textbf{1024}, 117338 (2026)

\bibitem{Asghar:2026kwd}
Z.~Asghar, M.~F.~Shamir, F.~Mofarreh, J.~Rayimbaev, O.~Sirajiddin and F.~Shayimov,
Nucl. Phys. B \textbf{1024}, 117338 (2026)


\bibitem{Asghar:2026que}
Z.~Asghar, M.~F.~Shamir, F.~Mofarreh, J.~Rayimbaev, O.~Sirajiddin and F.~Shayimov,
Nucl. Phys. B \textbf{1025}, 117417 (2026)


\bibitem{LIGOScientific:2017ync}
B.~P.~Abbott \textit{et al.} [LIGO Scientific and Virgo],
Astrophys. J. Lett. \textbf{848}, no.2, L12 (2017)


\bibitem{Guo:2023vzz}
L.~Guo and Y.~Niu,
Phys. Rev. C \textbf{110}, no.1, L012801 (2024)

\bibitem{Radice:2017lry}
D.~Radice, A.~Perego, F.~Zappa and S.~Bernuzzi,
Astrophys. J. Lett. \textbf{852}, no.2, L29 (2018)

\bibitem{Burgio:2018yix}
G.~F.~Burgio, A.~Drago, G.~Pagliara, H.~J.~Schulze and J.~B.~Wei,
Astrophys. J. \textbf{860}, no.2, 139 (2018)

\bibitem{Ruiz:2017due}
M.~Ruiz, S.~L.~Shapiro and A.~Tsokaros,
Phys. Rev. D \textbf{97}, no.2, 021501 (2018)

\bibitem{KAGRA:2021duu}
R.~Abbott \textit{et al.} [KAGRA, VIRGO and LIGO Scientific],
Phys. Rev. X \textbf{13}, no.1, 011048 (2023)


\bibitem{Bailyn:1997xt}
C.~D.~Bailyn, R.~K.~Jain, P.~Coppi and J.~A.~Orosz,
Astrophys. J. \textbf{499}, 367 (1998)

\bibitem{Ozel:2010su}
F.~Ozel, D.~Psaltis, R.~Narayan and J.~E.~McClintock,
Astrophys. J. \textbf{725}, 1918-1927 (2010)

\bibitem{Farr:2010tu}
W.~M.~Farr, N.~Sravan, A.~Cantrell, L.~Kreidberg, C.~D.~Bailyn, I.~Mandel and V.~Kalnera,
Astrophys. J. \textbf{741}, 103 (2011)

\bibitem{ArcaSedda:2021zmm}
M.~Arca Sedda,
Astrophys. J. Lett. \textbf{908}, no.2, L38 (2021)

\bibitem{Bhar2023} P. Bhar, A. Errehymy, S. Ray, Eur. Phys. J. C \textbf{83}, 1151 (2023)

\bibitem{Samsing:2019dtb}
J.~Samsing, D.~J.~D'Orazio, K.~Kremer, C.~L.~Rodriguez and A.~Askar,
Phys. Rev. D \textbf{101}, no.12, 123010 (2020)

\bibitem{Ye:2019xvf}
C.~S.~Ye, W.~f.~Fong, K.~Kremer, C.~L.~Rodriguez, S.~Chatterjee, G.~Fragione and F.~A.~Rasio,
Astrophys. J. Lett. \textbf{888}, no.1, L10 (2020)


\bibitem{Fragione:2020aki}
G.~Fragione, A.~Loeb and F.~A.~Rasio,
Astrophys. J. Lett. \textbf{895}, no.1, L15 (2020)

\bibitem{Lu:2020gfh}
W.~Lu, P.~Beniamini and C.~Bonnerot,
Mon. Not. Roy. Astron. Soc. \textbf{500}, no.2, 1817-1832 (2020)

\bibitem{Liu:2020gif}
B.~Liu and D.~Lai,
Mon. Not. Roy. Astron. Soc. \textbf{502}, no.2, 2049-2064 (2021)


\bibitem{Bartos:2023lfu}
I.~Bartos, S.~Rosswog, V.~Gayathri, M.~C.~Miller, D.~Veske and S.~Marka,
[arXiv:2302.10350 [astro-ph.HE]].


\bibitem{Farrow:2019xnc}
N.~Farrow, X.~J.~Zhu and E.~Thrane,
Astrophys. J. \textbf{876}, no.1, 18 (2019)

\bibitem{Kiziltan:2013oja}
B.~Kiziltan, A.~Kottas, M.~De Yoreo and S.~E.~Thorsett,
Astrophys. J. \textbf{778}, 66 (2013)

\bibitem{Valentim:2011vs}
R.~Valentim, E.~Rangel and J.~E.~Horvath,
Mon. Not. Roy. Astron. Soc. \textbf{414}, 1427 (2011)

\bibitem{Zhang:2010qr}
C.~M.~Zhang, J.~Wang, Y.~H.~Zhao, H.~X.~Yin, L.~M.~Song, D.~P.~Menezes, D.~T.~Wickramasinghe, L.~Ferrario and P.~Chardonnet,
Astron. Astrophys. \textbf{527}, A83 (2011)


\bibitem{Ovalle:2017fgl}
J.~Ovalle,
Phys. Rev. D \textbf{95}, no.10, 104019 (2017)

\bibitem{Ovalle:2018gic}
J.~Ovalle,
Phys. Lett. B \textbf{788}, 213-218 (2019)

\bibitem{daRocha:2020jdj}
R.~da Rocha,
Phys. Rev. D \textbf{102}, no.2, 024011 (2020)

\bibitem{Ovalle:2017wqi}
J.~Ovalle, R.~Casadio, R.~da Rocha and A.~Sotomayor,
Eur. Phys. J. C \textbf{78}, no.2, 122 (2018)

\bibitem{Ovalle:2018umz}
J.~Ovalle, R.~Casadio, R.~d.~Rocha, A.~Sotomayor and Z.~Stuchlik,
Eur. Phys. J. C \textbf{78}, no.11, 960 (2018)

\bibitem{Ovalle:2019lbs}
J.~Ovalle, C.~Posada and Z.~Stuchl\'\i{}k,
Class. Quant. Grav. \textbf{36}, no.20, 205010 (2019)

\bibitem{Casadio:2019usg}
R.~Casadio, E.~Contreras, J.~Ovalle, A.~Sotomayor and Z.~Stuchlick,
Eur. Phys. J. C \textbf{79}, no.10, 826 (2019)

\bibitem{Zubair:2020lna}
M.~Zubair and H.~Azmat,
Annals Phys. \textbf{420}, 168248 (2020)

\bibitem{Ovalle:2020kpd}
J.~Ovalle, R.~Casadio, E.~Contreras and A.~Sotomayor,
Phys. Dark Univ. \textbf{31}, 100744 (2021)

\bibitem{Ovalle:2021jzf}
J.~Ovalle, E.~Contreras and Z.~Stuchlik,
Phys. Rev. D \textbf{103}, no.8, 084016 (2021)

\bibitem{Contreras:2021yxe}
E.~Contreras, J.~Ovalle and R.~Casadio,
Phys. Rev. D \textbf{103}, no.4, 044020 (2021)



\bibitem{Maurya:2022cyv}
S.~K.~Maurya, A.~Errehymy, R.~Nag and M.~Daoud,
Fortsch. Phys. \textbf{70}, no.5, 2200041 (2022)

\bibitem{Maurya:2022uqu}
S.~K.~Maurya, M.~Govender, K.~N.~Singh and R.~Nag,
Eur. Phys. J. C \textbf{82}, no.1, 49 (2022)

\bibitem{Maurya:2022brt}
S.~K.~Maurya, A.~Banerjee, A.~Pradhan and D.~Yadav,
Eur. Phys. J. C \textbf{82}, no.6, 552 (2022)

\bibitem{Sharp-Misner1964} C. W. Misner, D. H. Sharp, Phys. Rev. 136, B571 (1964).

\bibitem{Gokhroo1994}M.K. Gokhroo, A.L. Mehra, Gen. Relativ. Gravit.26, 75 (1994)
\bibitem{TOV1} R.C. Tolman,  Phys. Rev., 55, 364 (1939).
\bibitem{TOV2} J.R. Oppenheimer and  G.M. Volkoff, Phys. Rev., 55, 374 (1939).
\bibitem{Harko:2002pxr}
T.~Harko and M.~K.~Mak,
Chin. J. Astron. Astrophys. \textbf{2}, no.3, 248 (2002)

\bibitem{reg} T. Regge, J.A. Wheeler, Phys. Rev. 108, 1063 (1957) 
\bibitem{thor} K.S. Thorne, A. Campolattaro,  Astrophys. J. 149, 591 (1967)
\bibitem{hin} T. Hinderer, Astrophys. J. 677, 1216-1220 (2008)

\bibitem{Hinderer:2009ca}
T.~Hinderer, B.~D.~Lackey, R.~N.~Lang and J.~S.~Read,
Phys. Rev. D \textbf{81}, 123016 (2010)


\bibitem{Chodos:1974}A. Chodos, R.L.Jaffe, K. Johnson, C.B.Thorn,\& V.F. Weisskopf, Phys. Rev. D 9, 3471 (1974)

\bibitem{PSRJ1810+1744}
R.~W.~Romani, D.~Kandel, A.~V.~Filippenko, T.~G.~Brink and W.~Zheng,
Astrophys. J. Lett. \textbf{908}, L46 (2021)

\bibitem{starPSR}
D. Kandel, R. W. Romani, ApJ, \textbf{892}, 101 (2020)



\bibitem{bondi1992anisotropic}  Bondi, H. 1992, MNRAS, 259, 365.

\bibitem{heintzmann1975neutron}  Heintzmann, H., \& Heintzmann, W. 1975, A\&A, 38, 51

\bibitem{chan1993dynamical}  Chan, R.,   Herrera, L.,  \& Santos, N. O. 1993, MNRAS, 265, 533

\bibitem{chan1992dynamical}  Chan, R.,  Herrera, L.,   \& Santos, N. O. 1992, CQGra, 9, 133

\bibitem{Moustakidis:2016ndw}
C.~C.~Moustakidis,
Gen. Rel. Grav. \textbf{49}, no.5, 68 (2017)

\bibitem{Koliogiannis:2018hoh}
P.~S.~Koliogiannis and C.~C.~Moustakidis,
Astrophys. Space Sci. \textbf{364}, no.3, 52 (2019)

\bibitem{ZHN1} B.K. Harrison, K.S. Thorne, M. Wakano, J.A. Wheeler, Gravitational Theory and Gravitational Collapse, University of Chicago Press (1965)

\bibitem{ZHN2} Ya.B. Zeldovich, I.D. Novikov, Relativistic Astrophysics Stars and Relativity, Vol. 1, University of Chicago Press (1971)



\end{thebibliography}
\end{document}